\documentclass[twocolumn,showpacs,superscriptaddress,secnumarabic,amsmath,nobibnotes,bm,pre,prb,nofootinbib]{revtex4-2}
\usepackage{graphicx}
\usepackage{dcolumn}
\usepackage{bm}
\usepackage{color}
\usepackage{hyperref}
\usepackage{tabularx}
\usepackage{amsmath}
\usepackage{siunitx}
\usepackage{comment}
\usepackage{xcolor}
\usepackage{multirow}
\usepackage{adjustbox}
\usepackage{tcolorbox} 
\usepackage{booktabs}  
\usepackage{array}    
\usepackage{colortbl} 
\usepackage{bm}

\begin{document}
\author{Davide Baiocco}
\email{davide.baiocco@phd.unipi.it}
\affiliation{Dipartimento di Fisica dell'Universit\`a di Pisa, Largo Bruno Pontecorvo 3, I-56127 Pisa,~Italy}
\author{Damiano Marian}
\email{damiano.marian@unipi.it}
\affiliation{Dipartimento di Fisica dell'Universit\`a di Pisa, Largo Bruno Pontecorvo 3, I-56127 Pisa,~Italy}
\author{Giulio Marino}
\email{giulio.marino@phd.unipi.it}
\affiliation{Dipartimento di Fisica dell'Universit\`a di Pisa, Largo Bruno Pontecorvo 3, I-56127 Pisa,~Italy}
\affiliation{INFN, Sezione di Pisa, Largo Bruno Pontecorvo 3, I-56127 Pisa, Italy} 
\author{\\ Paolo Panci}
\affiliation{Dipartimento di Fisica dell'Universit\`a di Pisa, Largo Bruno Pontecorvo 3, I-56127 Pisa,~Italy}	\affiliation{INFN, Sezione di Pisa, Largo Bruno Pontecorvo 3, I-56127 Pisa, Italy} 
\author{Marco Polini}
\affiliation{Dipartimento di Fisica dell'Universit\`a di Pisa, Largo Bruno Pontecorvo 3, I-56127 Pisa,~Italy}
\author{Alessandro Tredicucci}
\affiliation{Dipartimento di Fisica dell'Universit\`a di Pisa, Largo Bruno Pontecorvo 3, I-56127 Pisa,~Italy}

\bibliographystyle{apsrev4-1}

\title{Direct Dark Matter searches with Metal Halide Perovskites}	
\begin{abstract}
Polar materials with optical phonons in the meV range are excellent candidates for both dark matter direct detection (via dark photon-mediated scattering) and light dark matter absorption. In this study, we propose, for the first time, the metal halide perovskites MAPbI$_3$, MAPbCl$_3$, and CsPbI$_3$ for these purposes. Our findings reveal that CsPbI$_3$ is the best material, significantly improving exclusion limits compared to other polar materials. For scattering, CsPbI$_3$ can probe dark matter masses down to the keV range. For absorption, it enhances sensitivity to detect dark photon masses below $\sim 10~{\rm meV}$. 
The only material which has so far been investigated and that could provide competitive bounds is CsI, which, however, is challenging to grow in kilogram-scale sizes due to its considerably lower stability compared to CsPbI$_3$. Moreover, CsI is isotropic while the anisotropic structure of CsPbI$_3$ enables daily modulation analysis, showing that a significant percentage of daily modulation exceeding 1\% is achievable for dark matter masses below $40~{\rm keV}$.

\end{abstract}

\maketitle
	
 \section{Introduction}
  
The search for cold Dark Matter (DM) made of new particles hinges on detecting their extremely weak interactions with the Standard Model (SM). Detecting these faint signals in the laboratory demands highly sensitive experimental techniques. Traditional approaches, such as those based on nuclear recoil (e.g., Xenon1T~\cite{PhysRevLett.123.251801}) have set the standard for direct detection, steadily increasing sensitivity and narrowing the parameter space. However, these methods face inherent limitations in detecting DM masses below the nuclear scale, $m_N \simeq 1\,$GeV. While this limitation is not problematic for conventional searches targeting thermally produced heavy DM via the freeze-out mechanism, it becomes significant in alternative scenarios. Production mechanisms like freeze-in~\cite{Hall_2010,Bernal_2017}, or models involving hidden sector~\cite{hidden1,hidden2,hidden3}, may suggest the existence of lighter DM candidates. These possibilities necessitate innovative detection strategies beyond nuclear recoil to explore this largely unexplored parameter space. Thus, new experimental approaches are emerging to probe a wide range of DM models, particularly those targeting the keV–MeV DM mass range. 

With detector thresholds now reaching the $\sim 100~{\rm meV}$ scale and expected to improve further, single-phonon excitations present a promising pathway for DM detection~\cite{Singleph1,Singleph2,Singleph3,Singleph4,Singleph5,ZurekPRD2018,epsELF2,Singleph8,Singleph9,Singleph10,Singleph11,Singleph12,Singleph13,Trickle_2020}. To be sensitive to such light DM, the candidate target must have a sufficiently small excitations gap, as well as favorable kinematics for DM scattering. In this regards, polar materials have emerged as a promising target for future sub-MeV DM direct detection experiments. In particular, prior research~\cite{ZurekPRD2018} has focused attention on DM-phonon interaction in polar materials for the following reasons. First, these materials feature gapped optical phonons, that can be thought of as oscillating dipoles which have a sizable coupling to kinetically mixed dark photons. Second, these optical phonons are gapped excitations with typical energies ranging from few up to $\sim 100~{\rm meV}$. This is kinematically favorable for sub-MeV DM, allowing $\gg \si{meV}$ energy depositions with low momentum transfer. Third, being these materials typically semiconductors/insulators, suppression from screening effects is much smaller than in other materials such as superconductors~\cite{PAOLUCCISUPERCONDUCTING,PaolucciTES_DM_2020} and Dirac materials~\cite{DiracMat_PhysRevD.101.055005,DirectionalPRDZurek}.  Fourth, if the crystal is not isotropic, then this induces a directional dependence in the DM scattering~\cite{TargCompSinglph}.

\medskip
In this work, we propose that metal halide perovskites, a class of polar materials that has recently garnered significant attention for THz-based applications, hold promise as potential targets for future light DM experiments. These perovskites follow the general chemical formula $\text{ABX}_3$, where A denotes the cation, which can be organic (e.g., methylammonium (MA) or formamidinium (FA)) or inorganic (e.g., Cs), B represents the metal cation, typically Pb or Sn, and X corresponds to halide ions such as Cl, Br, or I. Research on metal halide perovskites is of high technological relevance, as they have catalyzed a new era of ``perovskite photovoltaics", achieving certified power conversion efficiencies surpassing 20\% within just a few years of rapid development~\cite{cite-key}. This positions them as highly promising materials for optoelectronic applications.
Among perovskite-based solar cell candidates, the most prominent are~\emph{hybrid} perovskites, which have an organic-inorganic composition (e.g., MAPbI$_3$), and~\emph{all-inorganic} perovskites such as CsPbI$_3$. Their electronic bandgaps typically range from around $1.6$-$1.7~{\rm eV}$ (for MAPbI$_3$ and CsPbI$_3$) to about $3.0~{\rm eV}$ (for MAPbCl$_3$). Regarding their vibrational properties, these materials exhibit optical phonons at energies of a few meV ($\sim 1~{\rm THz}$). These combined characteristics make them particularly well-suited for probing light DM  with small masses (on the order of $\sim$ keV), as their low-frequency optical phonons can facilitate detection. Additionally, their crystal structures are {\it anisotropic} allowing daily modulation analysis~\cite{TargCompSinglph,dailyMod1,dailyMod2}. Although both organic and all-inorganic materials exhibit outstanding reach for light DM detection, it has been observed~\cite{MAPIdegradation} that hybrid perovskites undergo a more pronounced degradation process, which can be attributed to the presence of the hydrophobic organic cation, e.g.~methylammonium, which can accelerate the degradation path~\cite{MAPIdegradation}, while all-inorganic perovskites do not, paving the way for enhanced device lifetimes~\cite{CsPbI3-lifetime}. 

This article is organized as follows. In Sec.~\ref{sec:sec2}, we review DM scattering mediated by a dark photon, outlining the key assumptions and developing approximations to identify material characteristics that enhance the detection rate. In Sec.~\ref{sec:methalper} we then focus on the properties of the materials under study and their projected reach in direct searches. Specifically, we demonstrate that, under the most simplified approximation, metal halide perovskites perform comparably to CsI, the best material currently proposed in the literature. In Sec.~\ref{sec:DFTanalysis} we then extend this analysis via standard {\it ab-initio} condensed matter modeling, which confirms our initial findings.  Finally in Sec.~\ref{sec:DailyModulation}, leveraging the anisotropic structures of metal halide perovskites, we perform a daily modulation analysis.
In Sec.~\ref{sec:sec3}, we analyze the reach of these materials in the context of direct dark photon absorption. In Sec.~\ref{sec:sec4}, we focus specifically on CsPbI$_3$, discussing in detail a possible experimental setup and its feasibility. Finally, in Sec.~\ref{sec:sec5}, we draw our main conclusions.

\section{Dark photon mediated scattering}
\label{sec:sec2}

Phonons produced via scattering off DM has gained attention as a promising method, particularly due to its potential sensitivity to sub-MeV DM particles. In general, a DM particle with mass \(m_\chi\) interacts with a target system at a scattering rate
\begin{equation}
\begin{split}
 R&= \frac{1}{\rho_{\rm T}}\frac{\rho_\chi}{m_\chi}\int d^3 \textbf{v} f_\chi(\textbf{v}) \Gamma(\textbf{v}) \ ,
 \label{rate}
 \end{split}
\end{equation}
expressed in terms of counts per unit of exposure.
Here, \( f_\chi(\textbf{v}) \) denotes the DM velocity distribution, assumed to follow the Standard Halo Model. The parameters for this model include an escape velocity of \( v_{esc} = 600~\si{km/s} \), a velocity dispersion of \( v_0 = 230~\si{km/s} \), and an Earth velocity of \( v_e = 240~\si{km/s} \)~\cite{LEWIN199687}. The local DM density, \( \rho_\chi = 0.4~\si{GeV}/\si{cm}^3 \)~\cite{Catena:2009mf, Salucci:2010qr}, while \( \rho_{\rm T} \) represents the density of the target material. In the case of an incoming DM particle with velocity \( \textbf{v} \) performing single phonon excitation, the general form of \( \Gamma(\textbf{v}) \) is given by \cite{Trickle_2020, epsELF2}:
\begin{equation}
    \Gamma (\mathbf{v})=\int \frac{d^3q}{(2\pi)^3}\mathcal{F}_{\rm med} ^2 (\mathbf{q})\sum_\nu 2\pi  \delta (\omega-\omega_{\nu,\mathbf{q}}) S_\nu(\mathbf{q}) \ ,
\end{equation}
where the sum is done over single phonon states labeled by branch $\nu$, momentum $\mathbf{k}$ and energy $\omega_{\nu,\mathbf{k}}$. Here the form factor simplifies to $\mathcal{F}_{\rm med}\propto (q_0/q)^2$ for a light mediator, and the DM-target response is controlled by the dynamic structure factor $S_\nu(\mathbf{q})$.

In this work, we investigate dark photon-mediated scattering, a promising mechanism for optical phonon excitation. We consider a fermionic dark matter interaction of the form $e' \bar{\chi} \gamma^\mu \chi A'_\mu$ where \(A'_\mu\) is a dark photon, associated with an extra $U(1)_{\rm X}$ symmetry, that kinetically mixes with the SM photon \(A_\mu\) via a mixing parameter \(\kappa/2\). In the limit of a mediator mass $m_{A'}\ll\si{eV}$, DM particles behave effectively as millicharged under the SM photon with coupling \( \kappa e' \bar{\chi} \gamma^\mu \chi A_\mu \) \cite{Knapen_2017}. The total DM-electron scattering cross section coming from the model under consideration is spin independent and for a fixed momentum transfer $q_0=\alpha m_e$ is given by
\begin{equation}
    \Bar{\sigma}_e\equiv \frac{16\pi \mu_{\chi e}^2\alpha\kappa^2 \alpha'}{q_0^4} \ ,
    \label{eq:sigmabare}
\end{equation}
where $\mu_{\chi,e}$ is the reduced mass of DM-electron system. Here, $\alpha$ is the electromagnetic fine structure constant and $\alpha'= e'^2/4\pi$. In the literature, $\Gamma(\mathbf{v})$ is often normalized to Eq.~\ref{eq:sigmabare}. 

\smallskip
It is worth to stress that, only polar materials with differently charged ions in the primitive cell, described by the Born effective charges, can couple phonon modes to the photon. The optical mode properties influence the shape of the phonon excitations curves, as analyzed in Ref.~\cite{epsELF2}. To compute the DM-phonon scattering rate, we draw an analogy with electron-phonon scattering, adjusting the electronic charge to reflect the effective DM charge. To this end, we make use of open-source codes like \texttt{PhonoDark} \cite{PhonoDark} and \texttt{DarkELF} \cite{DarkELF} for direct numerical calculations of the scattering rate for various material targets and DM models. \texttt{PhonoDark} uses first-principles methods based on Density Functional Theory (DFT) for electron/phonon excitations, while \texttt{DarkELF} relies on the energy loss function (ELF) of the material. 

In subsection~\ref{sec:ELFanalysis} we introduce the ELF formulation in order to show the key material parameters affecting the scattering rate, while in~\ref{sec:methalper} we focus on metal halide perowskites. In subsection~\ref{sec:DFTanalysis} we use the first-principles approach to gain deeper insights into the DM-phonon scattering process, including the material's anisotropy that affects the daily modulation signal.

\subsection{ELF analysis}
\label{sec:ELFanalysis}
The ELF is a quantity that can be measured experimentally and is defined as:
\begin{equation}
{\cal L}(\omega, \textbf{q}) = \text{Im}\left[ - \frac{1}{\epsilon(\omega,\textbf{q})}\right] \ ,
\end{equation}
where \( \epsilon(\omega,\textbf{q}) \) represents the dielectric function of the material, dependent on both momentum and frequency. In the sub-MeV DM mass regime, since $|\textbf{q}| \ll \text{keV}$, it is reasonable to take ELF $|\textbf{q}|$-independent. In the $q\to 0$ limit, the dynamic structure factor $S_\nu(\mathbf{q})\propto \bar{\sigma}_e\,\mathcal{L}(\omega,\mathbf{q})$ and therefore the scattering rate simplifies to:
\begin{equation}
\begin{split}
R & \approx \frac{1}{\rho_T}\frac{\rho_\chi}{m_\chi}\frac{\Bar{\sigma}_e}{\mu_{\chi e}^2}\frac{q_0^4}{4\pi \alpha}\int d^3 v \frac{f_\chi(v)}{v}\int \frac{d\omega}{2\pi} \\
& \quad \times \mathcal{L}(\omega,0)\log\bigg[\frac{1+\sqrt{1-2\omega/v^2m_\chi}}{1-\sqrt{1-2\omega/v^2m_\chi}}\bigg] \ .
\end{split}
\label{reach}
\end{equation}

For energies below the electronic band gap, the ELF is primarily influenced by energy losses associated with single phonon excitations. Hence, for these energies and in the absence of free carriers, the dielectric response arises from lattice absorption and can be expressed as:

\begin{equation}
\epsilon({\omega}) = \epsilon_{\infty} \prod_{\nu = 1}^{N} \frac{\omega^2_{\text{LO},\nu}-i\omega\gamma_{\text{LO},\nu}-\omega^2}{\omega^2_{\text{TO},\nu}-i\omega\gamma_{\text{TO},\nu}-\omega^2} \ ,
\label{eps_multi}
\end{equation}
where \( \epsilon_{\infty} \) accounts for the contributions from the vacuum and electronic interband transitions to the low frequency dielectric constant \cite{Gervais1974,ZollerJVST2019}. The quantities 
$\omega_{\text{TO/LO}}$ and $\gamma_{\text{TO/LO}}$ represent the transversal(T)/logitudinal(L) optical(O) phonons frequencies and dampings of the material, respectively. We will come back on these quantities in the following section. The product in Eq.~(\ref{eps_multi}) runs over all infrared-active phonon modes ($N$) in the material, this is typically a smaller number compared to the total number of atoms in the unit cell. It has been observed that, in materials with many phonon modes or with large TO/LO splittings, Eq.~(\ref{eps_multi}) often provides a better description of the infrared reflectance of insulators compared to the Drude–Lorentz model of independent oscillators \cite{ZollerJVST2019}.

We emphasize that this expression for the dielectric constant is strictly valid only under the isotropic approximation for the material. However, this approximation breaks down in the case of Al$_2$O$_3$ and the proposed perovskites, as we will exploit their anisotropy to study the material's daily modulation. For these materials, the correct expression will be described by a tensorial dielectric function. Nevertheless, the anisotropic nature impacts only the very low mass regime, introducing a deviation of at most a factor of $\sim2$. Therefore, employing the isotropic approximation proves useful in identifying the key features that enable us to probe even lower mass ranges. We would like to stress that the relation between the dielectric responds and ELF is broken if the mediator couples to charge fluctuations differently from the SM photon~\cite{ZurekPRD2018,epsELF2,Trickle_2020,TargCompSinglph,Singleph9}.

It is well known, as reported in Ref. \cite{Trickle_2020}, that in the narrow phonon width approximation, i.e. $\gamma_{\nu} \ll \omega_{\nu}$, and for a material that has a single optical phonon, the ELF can be directly connected with the coupling constant in the Fr\"ohlich Hamiltonian \cite{Frohlich}
\begin{equation}		\mathcal{H}_I=i \sqrt{4\pi \alpha'} \, C_F\sum_{\mathbf{k},\mathbf{q}}\frac{1}{|\mathbf{q}|}\bigg[c^\dagger_{\mathbf{q}}a^\dagger_{\mathbf{k}-\mathbf{q}}a_{\mathbf{k}}-c.c.\bigg] \ , 
 \label{frohHam}
	\end{equation}
	where $c^\dagger_{\mathbf{q}}$ and $a^\dagger_{\mathbf{k}}$ are phonon and $\chi$ creation operators respectively. Here, $\alpha'$ is the dark fine structure constant defined in Eq.~\ref{eq:sigmabare} and the coupling 
  \begin{equation}
  C_F=\bigg[\frac{\omega_{\rm LO}}{2V_{\text{cell}}}\bigg(\frac{1}{\epsilon_\infty}-\frac{1}{\epsilon_0}\bigg)\bigg]^{1/2} \ ,
  \label{CF}
  \end{equation}
where $V_{\text{cell}}$ is the primitive cell volume. Even though multiphonon processes are expected to contribute to the scattering rate, if we restrict to the sub-MeV DM mass regime and assume single-phonon coupling, the matrix element squared from Eq.~\eqref{frohHam} turns out to be proportional to $V_{\rm cell}C_F^2$ and therefore the scattering rate takes the form:
 \begin{equation}
 R \propto \frac{\omega_{\text{LO}}}{2} \left( \frac{1}{\epsilon_{\infty}}-\frac{1}{\epsilon_0} \right) \log\bigg[\frac{1+\sqrt{1-2\omega_{\text{LO}}/v^2m_\chi}}{1-\sqrt{1-2\omega_{\text{LO}}/v^2m_\chi}}\bigg] \ .
 \label{R_1ph}
 \end{equation}
 
In summary, the assumptions made to identify the key criteria for selecting optimal materials are as follows: \textit{i)} DM-phonon scattering is mediated via a light dark photon, \textit{ii)} isotropic dielectric response, \textit{iii)} transferred momentum $|\mathbf{q}| \ll \si{keV}$, with an energy loss function that is $|\mathbf{q}|$-independent, \textit{iv)} narrow phonon width approximation, with the condition $\gamma_\nu \ll \omega_\nu$, consistent with the Fröhlich Hamiltonian. Even though we focus on the low mass regime, it is important to emphasize that for higher DM masses, a quality factor $Q$ can be introduced to quantify the suitability of the target~\cite{epsELF2}. 

\smallskip
In the low DM mass regime, Eq.~(\ref{R_1ph}) provides straightforward guidelines for material selection aimed at: $i)$ enhancing the detection rate, and/or $ii)$ identifying materials capable of probing sub-MeV DM masses. Regarding the first point, it is clear that by fixing $\omega_{\text{LO}}$ and $\epsilon_\infty$, an enhancement of $\epsilon_0$ maximizes the rate.
This occurs in so-called quantum paraelectric materials, such as strontium titanium oxygen SrTiO$_3$ at low temperature~\cite{PhysRevB.19.3593}. We plan to explore this aspect in a future work. 
For the second point, reducing $\omega_{\text{LO}}$ allows the detection of DM masses in the 1--100 keV range. In this work, we focus on this latter aspect and demonstrate that:
\begin{itemize}
    \item[$\diamond$] Metal halide perovskites enable the detection of DM masses in the keV range;
    \item[$\diamond$] The projected detection rate is competitive with materials previously studied in the literature;
    \item[$\diamond$] The anisotropy of the material allows for daily modulation of the signal.
\end{itemize}

\subsection{Optical Phonons in Metal Halide Perovskites}
\label{sec:methalper}

Before presenting our main results, we briefly review phonons in crystals. In a material with $N$ atoms in the primitive cell, there are 3 acoustic phonons and $N-3$ optical phonons. For example, a material with two atoms in the unit cell has three acoustic phonons, two transverse acoustic (TA) and one longitudinal acoustic (LA), and three gapped optical phonons, two transverse (TO) and one longitudinal (LO). Acoustic modes display a linear dispersion for $q\rightarrow 0$ while the optical modes have non zero frequencies $\omega_{\text{LO/TO}}$ in the same limit. Acoustic phonons represent in-phase ion motion, whereas optical phonons out-of-phase motion for $q \rightarrow 0$. While optical phonons are essential for coupling to millicharged DM particles, this coupling does not occur if the atoms in the unit cell are identical, as in Ge or Si, where no net polarization arises from optical phonon oscillations. In polar materials, ions have nonzero Born effective charges $Z^*$, e.g. $Z^* = \pm2.27$ in GaAs, due to the polar GaAs bond. The out-of-phase motion of optical phonons generates oscillating dipole moments and long-range dipole fields, enabling optical phonon coupling to charged particles, including electrons and DM interacting through an ultralight dark photon mediator. In the latter case, DM behaves as if it carries a tiny electric charge.

\smallskip
Now, we focus our attention on three materials: two hybrid perovskites, i.e. MaPbBr$_3$ and MaPbI$_3$, and one all-inorganic perovskite CsPbI$_3$. Firstly, in Table \ref{tab:omegas} we summarize the TO and LO phonon frequencies and dampings of all three materials as extracted experimentally \cite{C6MH00275G,GammaCsPbI3_Lattice2_MatPhyToday_2023}.

\begin{table}[t!]
    \centering
    \renewcommand{\arraystretch}{1.1} 
    \setlength{\tabcolsep}{3pt} 
    \definecolor{pastelblue}{RGB}{173, 216, 230} 
    \definecolor{pastelgreen}{RGB}{204, 255, 204} 
    \begin{adjustbox}{width=\columnwidth,center}
    \begin{tabular}{@{\hskip 5pt} c @{\hskip 5pt}||@{\hskip 1pt} c @{\hskip 1pt} c @{\hskip 1pt} c @{\hskip 1pt} c @{\hskip 1pt} c @{\hskip 1pt}}
         \hline
    \toprule
    \multicolumn{1}{c}{$\cellcolor{pastelgreen}$ Crystal} 
    & \multicolumn{1}{c}{$\cellcolor{pastelblue} \epsilon_{\infty}$}
     & \multicolumn{1}{c}{$\cellcolor{pastelblue}  \omega_{\text{TO}}[\si{meV}]$}
     & \multicolumn{1}{c}{$\cellcolor{pastelblue}  \gamma_{\text{TO}}[\si{meV}]$}
     & \multicolumn{1}{c}{$\cellcolor{pastelblue}  \omega_{\text{LO}}[\si{meV}]$}
     & \multicolumn{1}{c}{$\cellcolor{pastelblue}  \gamma_{\text{LO}}[\si{meV}]$}
     \\ 
        \bottomrule
        \hline
        \hline
    \multirow{2}{*}{ MAPbI$_3$}
    & \multirow{2}{*}{5.0}  & 3.84 & 1.08 & 4.80 & 1.32\\
    &                       & 7.56 & 2.40 & 15.96 & 2.40 \\
     \hline 
         \hline 
    \multirow{2}{*}{MAPbBr$_3$}
    & \multirow{2}{*}{4.7} & 5.40 & 1.20 & 6.12 & 1.80\\
    &                      & 8.76 & 3.60 & 20.04 & 3.24 \\
    \hline 
        \hline 
    \multirow{6}{*}{ $\gamma$-CsPbI$_3$}
    & \multirow{6}{*}{5.0} & 3.72 & 0.83 & 3.87 & 0.84 \\
    &                      & 5.79 & 1.24 & 6.71 & 1.22 \\
    &                      & 7.44 & 0.83 & 7.78 & 1.09 \\
    &                      & 8.68 & 1.65 & 9.38 & 1.89 \\
    &                      & 10.34 & 0.83 & 10.36 & 0.89 \\
    &                      & 11.17 & 2.89 & 17.55 & 2.34 \\
              \hline
              \bottomrule
              \hline
     \end{tabular}
     \end{adjustbox}
     
     \caption{TO and LO phonon frequencies, along with their corresponding damping factors for MAPbI$_3$, MAPbBr$_3$, and CsPbI$_3$. Data are taken from Refs.~\cite{C6MH00275G,GammaCsPbI3_Lattice2_MatPhyToday_2023}.}
          \label{tab:omegas}
     \end{table}

 \begin{figure}[t!]
    \centering
    \includegraphics[width=0.95\columnwidth]{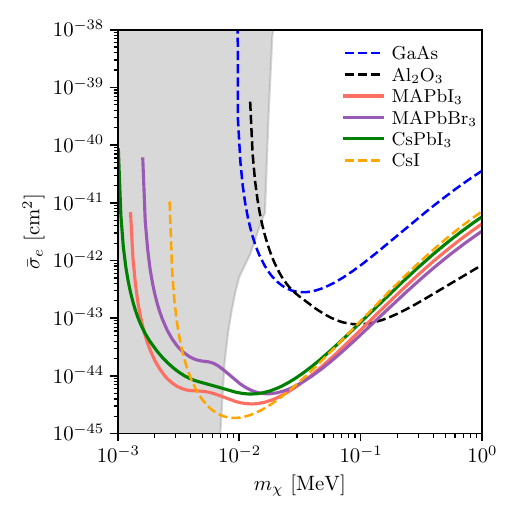}
    \caption{Reference DM-electron cross section as a function of $m_\chi$. The sensitivity is computed for a kg-year exposure on CsPbI$_3$, MAPbI$_3$, MAPbBr$_3$ (solid lines)  compared with other materials already analyzed in literature (dashed lines). We plot the cross sections needed to obtain 3 events for a kg-year exposure corresponding to a 95\% confidence level exclusion limit assuming a zero background. We used the parameters reported in Table \ref{tab:omegas} within the narrow width approximation. The light gray shadow region indicates stellar constraints~\cite{Stellar1}.}
    \label{fig:sigma}
\end{figure}

It can be observed that modeling the dielectric function of hybrid perovskites requires two infrared modes, while for $\gamma$-CsPbI$_3$, six modes are necessary due to its orthorhombic crystal structure. By simple inspection of the frequencies we see that all the materials present LO phonons at few meV. Until now, to the best of our knowledge, only CsI, between the materials investigated in the literature \cite{epsELF2}, presents comparable optical frequencies ($\omega_{\text{TO}}$ = 7.44 meV and $\omega_{\text{LO}}$ = 10.27 meV \cite{CsI_Frequencies}), while other materials that have been proposed and extensively studied for direct detection of light DM, such as GaAs and Al$_2$O$_3$, present optical frequencies greater than 30 meV.
Drawbacks of CsI are its hygroscopicity and its Mohs hardness in the 1-2 range, both reflecting in greater surface irregularities and scattering~\cite{kaplunov2021transmittance}. Moreover, it has a cubic crystal structure (space group Pm$\bar{3}$m, No. 221) that prevents the possibility of noise subtraction through daily modulation of the observed signal. 
 
As described in~\ref{sec:ELFanalysis}, we have calculated the rate using the narrow phonon width approximation. The projected 95\% confidence level (CL) exclusion reach on $\Bar{\sigma}_e$ is obtained by computing the cross section needed to obtained 3 events for a kg-year exposure for single-phonon excitation. We use the material parameters reported in Tab.~\ref{tab:omegas} and the results are shown in Fig.~\ref{fig:sigma}. 

Our findings highlight that metal halide perovskites substantially enhance sensitivity when compared to other materials widely explored in the literature. Specifically, the peak performance is achieved with \(\mathrm{MaPbI}_3\), reaching a sensitivity of $\Bar{\sigma}_e=4\cdot 10^{-45}\,\si{cm^2}$ at $m_\chi=14$ keV.

Moreover, they are competitive in terms of the maximum achievable detection rate, even when compared to CsI, which, to the best of our knowledge, offers the highest rate for light DM detection~\cite{epsELF2}. 
Additionally, the lower phonon gap in metal halide perovskites enables new kinematic possibilities, allowing for the exploration of the keV mass scale. This makes it feasible to probe the parameter space of a warm DM candidate through Earth-based experiments. In this context, it is important to highlight that stellar bounds provide complementary exclusions~\cite{Stellar1, Stellar2, Stellar3}. The latest limits are taken by~\cite{Stellar1} and represented by the light gray shaded region in Fig.~\ref{fig:sigma}, which excludes the mass range below $\sim7$ keV. Although these bounds are generally unavoidable, they might be evaded by introducing an ultra weak interaction between \( \chi \) and additional degrees of freedom. For example, an ultralight scalar field \( \phi \), with Yukawa couplings to both DM and SM nucleons, can induce an effective mass proportional to the vacuum expectation value of \( \phi \). In regions of high nucleon density, \( \chi \) acquires a significant mass, suppressing its thermal production in stellar cores and thereby relaxing all stellar cooling constraints~\cite{Stellar2}.

\smallskip
In summary, within the region allowed by stellar bounds, metal halide perovskites provide significantly more stringent limits compared to GaAs and Al$_2$O$_3$, raising important questions about their comparison with CsI in terms of scalability and the construction of next-generation detectors.

\subsection{DFT Analysis}
\label{sec:DFTanalysis}
In the following, we extend our analysis to predict the rate by carrying out {\it ab-initio} calculations, using the open-source code \texttt{PhonoDark} \cite{PhonoDark}. This requires the phonon dispersion of the material as input, which we obtaine through first-principle methods. To this end,  we make use of the Quantum Espresso suite \cite{QE-2009,QE-2017}. Because of the similar rates of the three perovskites, we focus our attention only on one of them, i.e.~CsPbI$_3$, which is an all-inorganic compound that is expected to suffer less from degradation compared to hybrid perovskites.

\begin{figure}[t!!!]
    \centering
    \includegraphics[width=0.45\columnwidth]{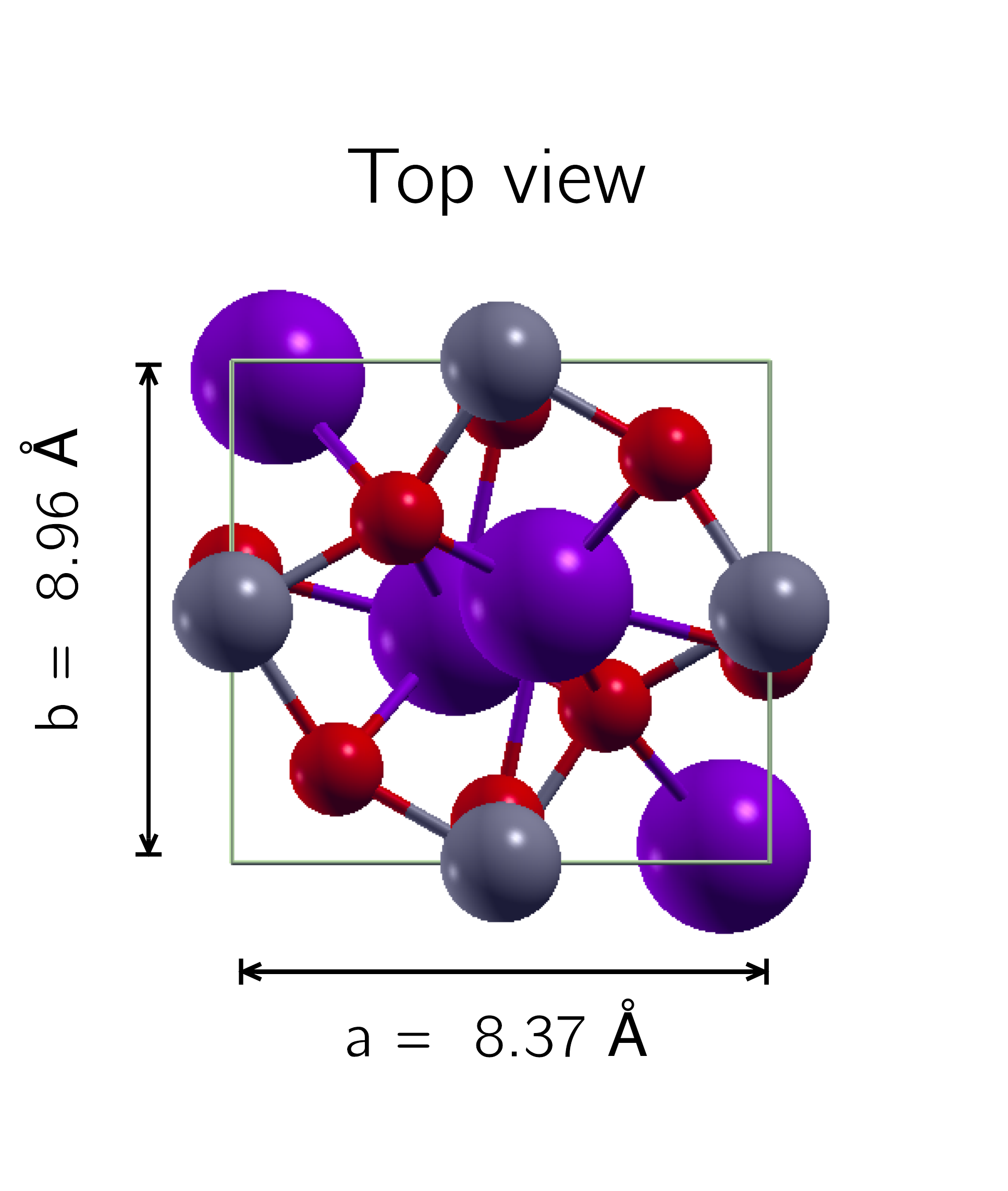}\hspace{0.3cm}
    \includegraphics[width=0.41\columnwidth]{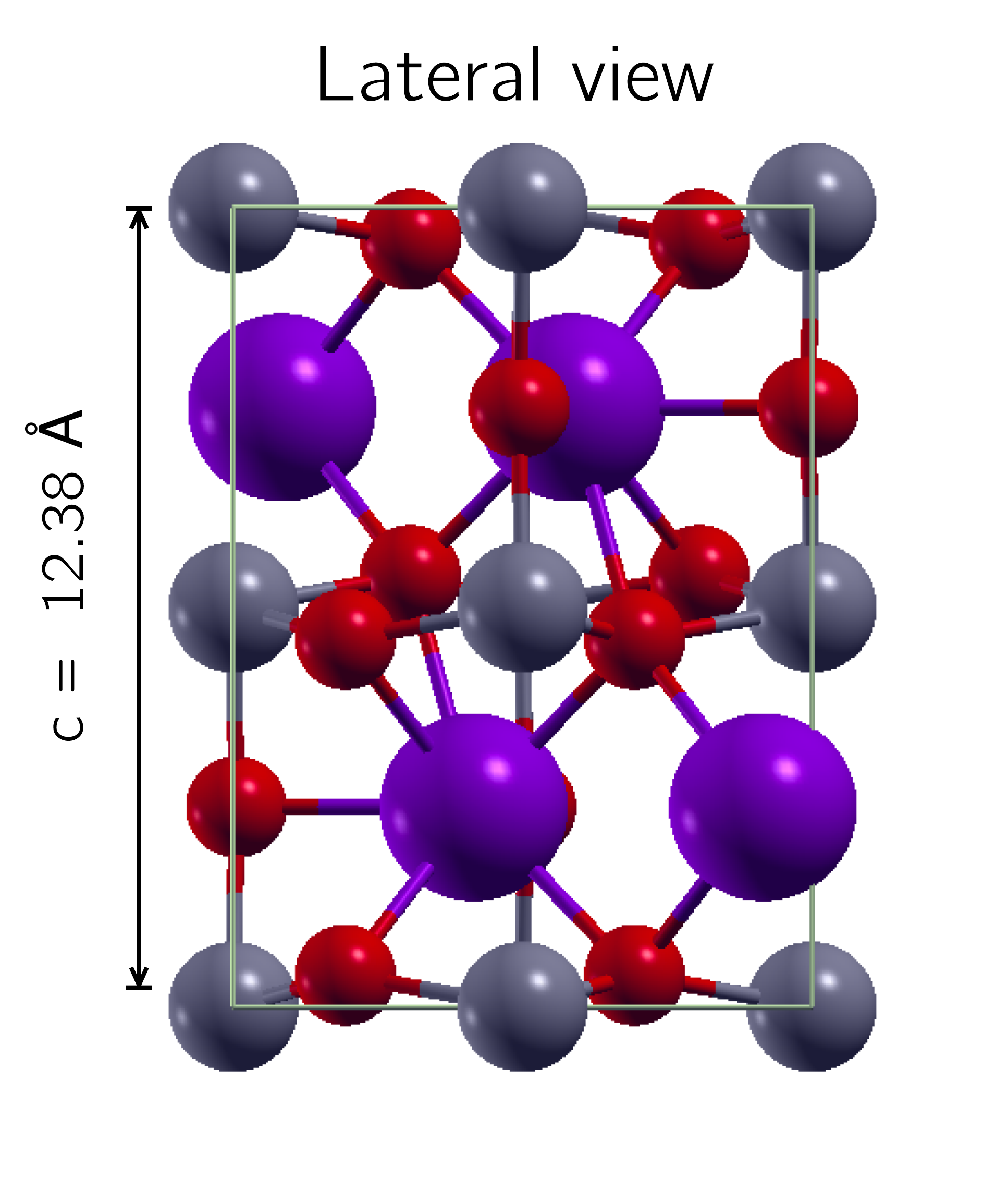}\\
    \vspace{0.1 cm}
    \includegraphics[width=0.9\columnwidth]{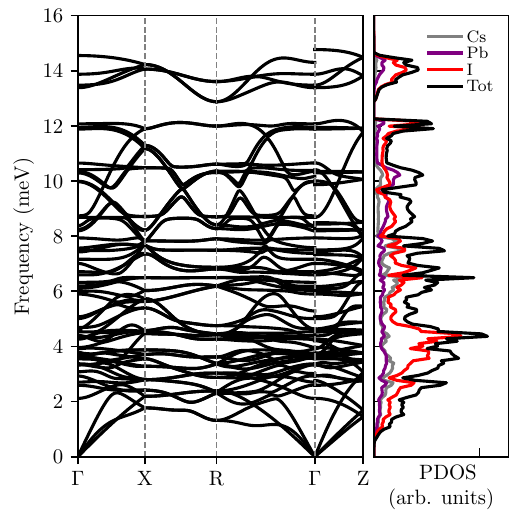}
    \caption{\textbf{Top panel:} Crystal structure of \(\gamma\)-CsPbI\(_3\) with lattice parameters, shown from both top and lateral views. \textbf{Bottom panel:} Phonon dispersion on the $\Gamma$-X-R-$\Gamma$-Z high symmetry path (left) and PDOS (right).}
    \label{fig:cspbi3}
\end{figure}

Among the various crystal structures that CsPbI$_3$ can assume~\cite{GammaCsPbI3_Marronier_ACSNano_2018}, we focused on the $\gamma$ structure, since it is the only kinetically stable perovskite-like structure~\cite{GAMMACSPBI3} (Pnma space group, GdFeO$_3$-like structure, tilt system \#10 according to Glazer~\cite{GLAZERPEROVSKITES}). The crystal structure of the $\gamma$-CsPbI$_3$ phase is schematically reported in the top panel of Fig.~\ref{fig:cspbi3}. Additionally, we analyze the thermodynamically stable structure, referred to as $\delta$-CsPbI$_3$. This structure does not belong to the perovskite family, but it shares the same orthorombic lattice and similar vibrational properties of the $\gamma$ structure, although with different lattice constants, and the space  group (Pmna). A more detailed description of $\delta$-CsPbI$_3$ is reported in the Appendix~\ref{app:app1}.

We perform structural relaxation until the forces for all the atoms are smaller than $10^{-4}$ Ry/\AA, sampling of the Brillouin zone with a 9x9x7~$\Gamma$-centered Monkhorst–Pack (MP) grid \cite{MPgrid}. Self-consistent calculation is performed on the same grid with convergence threshold of $10^{-12}$~Ry. All the calculation have been performed using projector-augmented wave pseudopotentials with PBEsol exchange-correlation functional with cutoffs of 65~and~360~Ry for the wave functions and the density respectively. The equilibrium lattice constants of $\gamma$-CsPbI$_3$ obtained after structural optimization are a~$=~8.37$~\AA, b~$=~8.96$~\AA, and c~$=~12.38$~\AA, which are in good agreement with previous reports \cite{GammaCsPbI3_Lattice2_MatPhyToday_2023}. Afterwards, we compute the phonon dispersion with the frozen method as implemented in \texttt{Phonopy} \cite{phonopy-phono3py-JPCM,phonopy-phono3py-JPSJ} using a 3x3x2 supercell on a 2x2x2~$\Gamma$-centered MP grid, and including non-analytical term correction \cite{nac1,nac2,nac3}. The phonon dispersion relation and the Projected Density of States (PDOS) for each atom species are reported in the bottom panel of the same figure. The calculation confirms that this material has very low, less than 1 THz, optical phonons and that the contribution is mainly due to the I-atoms coupling with Cs and Pb atoms, as can be inferred from the PDOS.  

Having the phonon dispersion relation allows us to directly compute the rate using the phonon dispersion obtained, see, e.g., Refs. \cite{Knapen_2018,Trickle_2020}, as implemented in the \texttt{PhonoDark} code \cite{PhonoDark}.
It is worth noting that results obtained using ab-initio methods and ELF have been compared in Ref. \cite{DarkELF}, showing good agreement. 
It is important to note that the ELF results are based on experimentally measured optical frequencies \cite{C6MH00275G,GammaCsPbI3_Lattice2_MatPhyToday_2023}, whereas the DFT method is not, which can lead to some discrepancies in the cross section. With this caveat in mind, we can use first-principle calculations to further investigate the properties of CsPbI$_3$ for direct DM detection. 

Conversely to the narrow phonon width approximation used in previous sections, here the dynamical structure factor $S(\mathbf{q},\omega)$ takes a more general form~\cite{Trickle_2020}:
\begin{equation}\label{eq:structure_factor_for_DFT}
    S_\nu(\mathbf{q})=\frac{1}{2 V_{\rm cell} \omega_{\nu,\mathbf{k}}}\bigg| \sum_j \frac{e^{-W_j(\mathbf{q})}}{\sqrt{m_j}}e^{i \mathbf{G}\cdot \mathbf{x}_j^0}(\mathbf{Y}_j\cdot \bm{\epsilon}^*_{\nu,\mathbf{k},j}) \bigg|^2 \ ,
\end{equation}
where $m_j$ and $\mathbf{x}_j^0$ are the masses and equilibrium positions of the ions, respectively. In Eq.~(\ref{eq:structure_factor_for_DFT}),  $\bm{\epsilon}_{\nu,\mathbf{k},j}$ and $\omega_{\nu,\mathbf{k}}$ are the phonon eigenvectors and energies, and $W_j(\mathbf{q})$ is the Debye-Weller factor. Notice that the interaction is governed by the model dependent DM-ion coupling $\mathbf{Y}_j$, with $j$ labeling the ions in the primitive cell, that in the case of a dark photon mediated scattering reduces to $\mathbf{Y}_j=-\mathbf{q}\cdot \mathbf{Z}_j^*/\hat{\mathbf{q}}\cdot \bm{\epsilon}_\infty\cdot \hat{\mathbf{q}}$, where $\mathbf{Z}_j^*$ is the Born effective charge tensor\footnote{
We fixed a typo in the \texttt{PhonoDark} code to properly account for the non-symmetric nature of the $Z^\star$ tensor.} of the $j^{\rm th}$ ion and $\bm{\epsilon}_\infty$ the high-frequency dielectric tensor that takes into account the electronic contribution to in-medium screening.

\begin{figure}[h!!!]
    \centering
    \includegraphics[width=0.9\columnwidth]{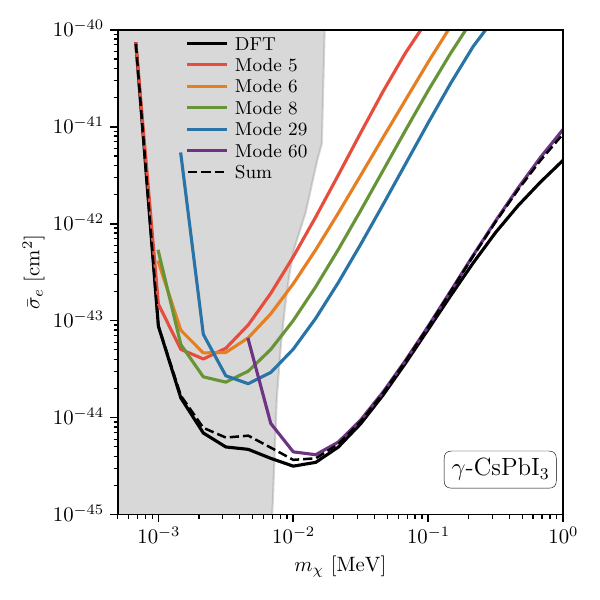}
    \vspace{0.cm}
    \includegraphics[width=0.9\columnwidth]{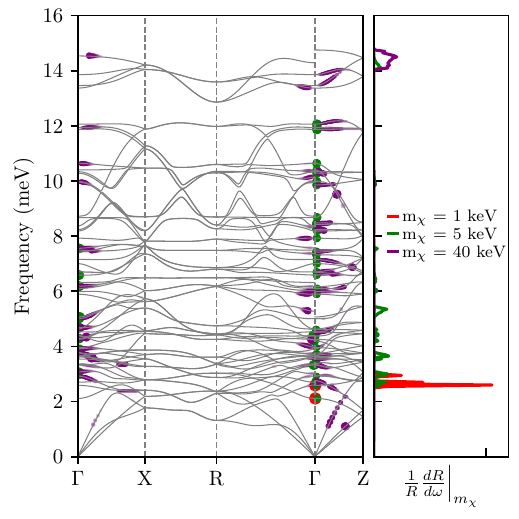}
    \caption{\textbf{Top panel:} $\Bar{\sigma}_e$ as a function of $m_\chi$ as in Fig.~\ref{fig:sigma}. Black solid line shows the total rate obtained through DFT, while with different colors denote the contribution of five modes. Dashed red curves the rate obtained summing these five modes. \textbf{Bottom panel:}  (Left) Rate contribution for three different $m_{\chi}$ masses, projected onto the phonon dispersion along the $\Gamma$-X-R-$\Gamma$-Z high symmetry path. (Right) Total rate Density of States for three different $m_{\chi}$ masses (1, 5 and 40 keV).  In the left panel, the point sizes are proportional to the normalized rate for each mass, considering only the contributions along the specified high-symmetry path. In the right panel, the differential rate $1/R\, dR/d\omega|_{m_\chi}$ represents the normalized rate contributions across the entire Brillouin zone for each mass.}
    \label{fig:ratemodes}
\end{figure}

At this point we can first isolate the contribution of each phonon mode to the total cross section. In the top panel of Fig.~\ref{fig:ratemodes}, we report the reference cross section $\Bar{\sigma}_e$ as a function of the DM mass, explicitly showing the contribution of four selected phonon modes and their sum. We can clearly observe that the total rate can be ascribed mainly to these four modes in almost the whole range of DM masses considered. It is also interesting to observe that the contribution to the lower masses can be entirely attributed to the lower longitudinal optical phonons of CsPbI$_3$, \textit{i.e.} modes with energies in the range $2-3$ meV, while for higher masses the rate can be ascribed mainly to the highest LO mode ($\omega_{\text{mode 60}} \sim 14.5~\si{meV}$), in accordance with Eq.~(\ref{R_1ph}). In the bottom panel of the same figure, we show the rate contributions for three different $m_{\chi}$ values, namely 1, 5, and 40 keV, projected onto the same high-symmetry k-path as in the bottom panel of Fig.~\ref{fig:cspbi3}. Additionally, we present the differential rate $R^{-1}\, dR/d\omega|_{m_\chi}$ integrated over the entire Brillouin zone for the same three $m_\chi$ values taken as a reference before. As a result, for lower DM masses of 1 keV and 5 keV, the rate contributions are predominantly concentrated at the $\Gamma$ point. In contrast, for a higher DM mass of 40 keV, the contributions extend to additional points within the Brillouin zone. Notably, the differential rate explicitly shows that for masses above 10 keV, only the highest optical mode contributes to the total reach, whereas for lower DM masses, multiple modes significantly contribute to the total rate.

\begin{figure*}[t!!!]
    \centering
        \includegraphics[width=0.975\columnwidth]{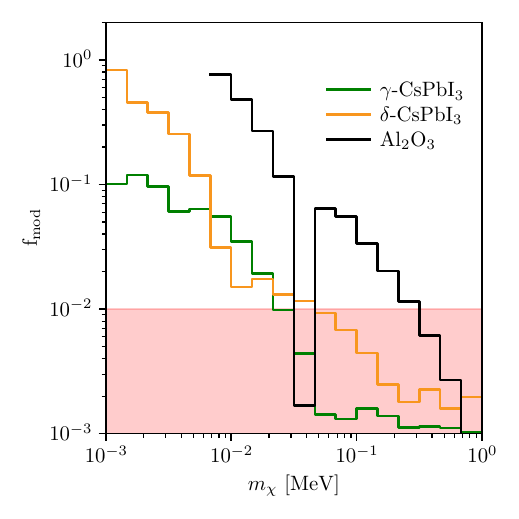} 
    \includegraphics[width=0.975\columnwidth]{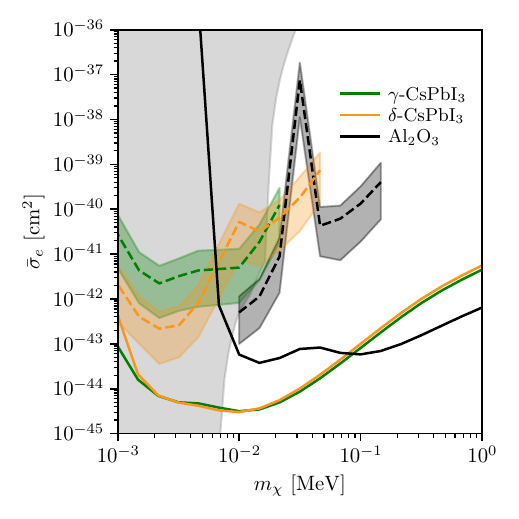} 
    \caption{\textbf{Left panel:} The maximal daily modulation amplitude, \( f_{\rm mod} \), as a function of \( m_\chi \), as defined in Eq.~\ref{eq:fmod}, is shown for CsPbI$_3$—with the green and orange solid lines representing the $\gamma$ and $\delta$ phases, respectively. The black solid line corresponds to the results for Al$_2$O$_3$ as computed in~\cite{ZurekPRD2018, TargCompSinglph}. The red shaded area indicates the region where \( f_{\rm mod} < 1\% \), highlighting that the statistical significance is too low to robustly confirm the modulating nature of the DM signal.
    \textbf{Right panel:} The cross-section $\Bar{\sigma}_e$ as a function of DM mass for $\gamma$-CsPbI$_3$ (solid green), $\delta$-CsPbI$_3$ (solid orange) and Al$_2$O$_3$ (solid black), calculated using DFT. The dashed lines represent the 95\% CL exclusion limits assuming zero observed events and no background. The $\pm1\sigma$ bands indicate the modulation reach for DM masses with $f_{\rm mod} > 0.01$, where the non-modulating hypothesis can be rejected and the statistical significance of a non-modulating signal can be established. Moreover the gray shaded region represents the stellar constraints as in Fig.~\ref{fig:sigma}.}
    \label{fig:daily}
\end{figure*}
\subsection{Daily Modulation}
\label{sec:DailyModulation}

\smallskip
Finally, we compute the daily modulation of the signal due to the anisotropy of the crystal exploiting the ab-initio computation. Indeed, the orientation of the detector changes with respect to the DM wind due to the rotation of the Earth around its own axis: if the crystal is anisotropic this results in a daily modulation of the rate. In addition, since the modulation pattern depends on the crystal orientation, running an experiment with multiple detectors simultaneously with different orientations can further enhance the signal-to-noise ratio. This effect has been studied firstly for Al$_2$O$_3$ \cite{ZurekPRD2018} and then for other materials \cite{DirectionalPRDZurek}.

Defining the maximal percentage deviation of the detection rate \( R \) from its daily average \( \langle R \rangle \) as a function of the DM mass
\begin{equation}
    f_{\rm mod} \equiv \frac{{\rm max}(|R - \langle R \rangle|)}{\langle R \rangle} \ ,
    \label{eq:fmod}
\end{equation}
one can identify the mass window allowed by imposing the condition \( f_{\rm mod} \gtrsim 1\% \), ensuring statistical relevance.

\smallskip
Left panel of Fig.~\ref{fig:daily} shows the daily modulation amplitude for CsPbI$_3$ (green and orange solid lines for both the $\gamma$ and $\delta$ phases) as a function of the DM mass assuming a detector energy threshold of $\omega_{\rm min} = 1\,\si{meV}$. For dark photon mediated scattering, the energy threshold does not affect significantly either the reach and the daily modulation amplitude, except at the lowest $m_\chi$ values. Black solid line represents the results for Al$_2$O$_3$ computed in~\cite{ZurekPRD2018, TargCompSinglph}. The dip arising in Al$_2$O$_3$ is due to a non trivial behavior of $f_{\rm mod}$. In particular, these dips correspond to peaks in the cross section, where it becomes possible to reject the non-modulating hypothesis~\cite{TargCompSinglph}. As one can see, $\gamma$-CsPbI$_3$ exhibits a significant modulation for DM masses around $\approx1$ keV, whereas modulation becomes statistically insignificant for masses $\gtrsim10$ keV. On the other hand, the $\delta$-CsPbI$_3$ allows achieving sensitivity to DM masses up to 40 keV, significantly covering the dip of Al$_2$O$_3$.

To evaluate the statistical significance of a modulating signal, we determine the expected number of events required to reject the non-modulating hypothesis, following the methodology described in~\cite{TargCompSinglph}. In the right panel of Fig.~\ref{fig:daily}, the black, green and orange solid lines represent the reference DM-electron cross-section \(\bar{\sigma}_e\) as a function of \(m_\chi\), calculated using DFT analysis for Al\(_2\)O\(_3\), $\gamma$-CsPbI\(_3\), $\delta$-CsPbI\(_3\), respectively. In the same figure, dashed lines show the modulating 95\% CL exclusion limits and the shaded regions the $\pm 1\sigma$ bands for which the non-modulating hypothesis can be rejected and the statistical significance of a non-modulating signal can be established, \textit{i.e.} $f_{\rm mod}>1\%$. 
As a result, $\gamma$-CsPbI$_3$ exhibits enhanced sensitivity to the modulating signal, but only for low DM masses. This material explores a region of parameter space that remains inaccessible to other materials; however, it is already subject to constraints from stellar bounds. Complementary to the behavior of the $\gamma$-phase, $\delta$-CsPbI$_3$, with its distinct crystal structure, can probe modulating hypotheses up to a DM mass of 40 keV, offering a competitive, if not superior, exclusion compared to Al$_2$O$_3$.

 \begin{figure}[t!]
    \centering
    \includegraphics[width=.95\columnwidth]{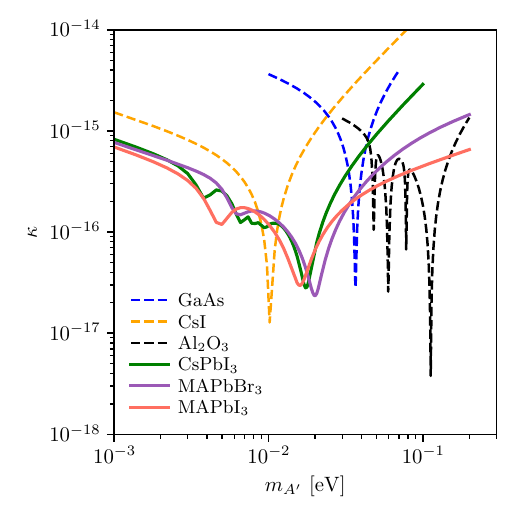}
    \caption{Exclusion limit at 95\% CL cross section reach with kg-yr exposure and zero background. The exclusion limits are computed using \texttt{DarkELF} \cite{DarkELF}, in terms of the kinetic mixing parameter $\kappa$ for kg-year exposure (single phonon excitation). 
    The width of the dips depends on both the parameter $\gamma$ and the temperature at which the ELF experimental data were collected.}
    \label{fig:Abs}
\end{figure}

 \section{Dark Photon Absorption}
 \label{sec:sec3}
In this section, we examine the absorption of dark photons into optical phonon excitations in polar materials. For sub-keV masses, dark photons are a viable DM candidate, detectable via an optical absorption signal if there is small mixing with the SM photon. Specifically, we consider DM consisting of non-thermally produced dark photons with kinetic mixing, given by $-\kappa F'_{\mu\nu}F^{\mu\nu}/2$. Polar materials are sensitive to dark photons in the mass range from a few meV up to hundreds of meV, due to the broad range of phonons that couple to electromagnetic fields \cite{ZurekPRD2018}. Since our focus is on this energy window, we neglect contributions from electronic excitations, which are sensitive to higher DM masses.

The absorption rate per unit target mass of a dark photon can be determined using the energy-loss function in the zero-momentum limit \cite{An_2015,Hochberg_2016}:

\begin{equation}
    R = \frac{\rho_\chi}{\rho_{\rm T}} \kappa^2  \, \text{Im}\left[-\frac{1}{\epsilon(m_{A'})}\right] \ ,
    \label{eq:abs}
\end{equation}
where $\kappa$ is the kinetic mixing parameter between the dark photon and the SM photon. The only dependence on the mass of the dark photon $m_{A'}$ comes directly from the ELF. Since optical measurements probe the zero-momentum limit of the dielectric function, previous studies have used this data to obtain the absorption rate. The absorption cross section is computed using \texttt{DarkELF}~\cite{DarkELF} using the parameters given in Tab.~\ref{tab:omegas}. 

\smallskip
Fig.~\ref{fig:Abs} shows the 95\% CL sensitivity on the $\kappa$ mixing parameter as a function of $m_{A'}$ for a kg-year exposure and zero background. As a result, metal halide perovskites (green, pink and red solid lines) exhibit significant sensitivity with respect to other targets (different dashed lines), providing the best limits for dark photon masses below $\sim10$ meV.
The obtained solid curves do not exhibit sharp resonance-induced behavior; instead, they appear smoother due to the broad nature of the several phonon modes in the 1-10 meV range. 
	
\section{Experimental Implementation}
\label{sec:sec4}

In this Section, we focus on CsPbI$_3$ discussing a possible experimental setup.

The first requirement to realize a DM detector would be the possibility to grow a kg-size CsPbI\textsubscript{3} single crystal.
The growth of $\gamma$ and $\delta$ single crystals has already been demonstrated either by Bridgman technique or by aqueous solution~\cite{BRIDGMANGROWTHCSPBI3,GAMMACSPBI3,CSPBI3AQUEOUS}. Both methods are in principle scalable to reach the desired size, as already achieved for different materials~\cite{BRIDGMANGROSSO}.

After polishing the crystal, one approach to reveal the presence of the phonons generated by the interaction of a single DM particle with the crystal could be fabricating an array of Kinetic Inductance Detectors (KID) on the crystal surface.
KIDs are superconducting resonators whose properties, in particular the kinetic inductance, strongly depends on the Cooper pair density in the superconducting material~\cite{DOYLEKID}.
The detector is commonly realized designing a mm-size LC circuit made of superconducting materials as Al or Ti-TiN systems~\cite{CALDER,BULLKID}.
The circuits are composed by an interdigitated capacitor connected to an inductive meander, acting as an oscillating circuit with quality factors of about 10\textsuperscript{4}-10\textsuperscript{5}.
The resonance frequency of this circuit can be approximated by~\cite{DOYLEKID}:
\begin{equation}
    \omega_0=\frac{1}{\sqrt{CL_{\rm TOT}}}=\frac{1}{\sqrt{C\left(L_{\rm KIN}+L_{m}\right)}} \ ,
\end{equation}
where $C$ is the capacitance, $L_{\rm KIN}$ is the kinetic contribution to the inductance and $L_{m}$ is the geometrical one. 

\begin{figure}[t!]
    \centering
\includegraphics[width=0.6\columnwidth]{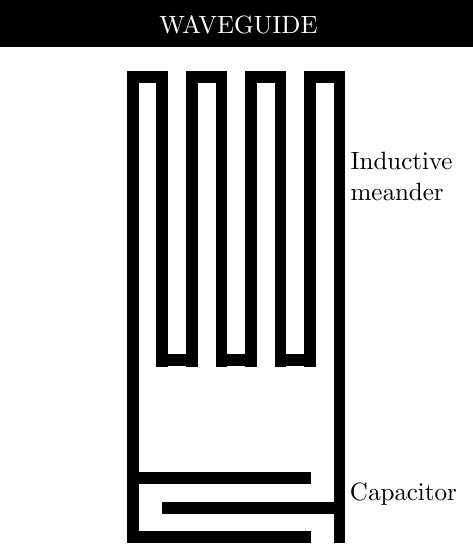}
    \caption{Schematic design of the LC superconductive resonator inductively coupled to a waveguide.}
    \label{fig:RES}
\end{figure}

The presence of phonons can break Cooper pairs inside the semiconductor, resulting in a variation of the kinetic inductance of the resonator and, thereby, the resonance frequency of the system:
\begin{equation}
    \delta f=-\frac{1}{2}f_0\zeta\frac{\delta L_{\rm KIN}}{L_{\rm KIN}} \ .
\end{equation}
where $\zeta=L_{\rm KIN}/L_{\rm TOT}$. In such devices, the value of $\zeta$ is of order 5-8\%~\cite{BULLKIDARRAY,CALDER}.
The resonators can be inductively or capacitively coupled to a coplanar waveguide and the transmission through the system of a RF signal can be measured to observe variation of the resonant frequency. 
A schematic of the resonator design is reported in Fig.~\ref{fig:RES}

This geometry allows the synchronous measurement of an array of resonators from the same waveguide by designing resonators with slightly different resonant frequencies and demultiplexing the different channels after the transmission in the waveguide.
By measuring amplitude variation of the transmitted signal at each frequency, it is possible to reveal the presence of phonons inside the substrate under each resonator.
Moreover, it is possible to measure also the phase shift that can improve the signal to noise ratio of the measurement.

Such devices have already been fabricated by lithography on silicon substrates and demonstrated the possibility of working as a photon detector~\cite{BULLKID}. In addition, the interaction between superconducting devices and acoustical phonon Fock states has already been observed~\cite{QUBITPHONONS}. The lithographic techniques for the fabrication of the above devices can be easily adapted to CsPbI$_3$ substrates, but would be challenging on less stable materials like CsI.
Moreover, it could be possible to fabricate an array of detectors on smaller crystals with different orientations to perform a parallel analysis of the daily modulation.

To detect the presence of a few phonons generated by the interaction with DM, it is necessary to suppress the thermal population of phonons by working at low temperature.
To estimate the working temperature, we calculated the population of phonons as:
\begin{equation}
    \bar{n}=\int_{\frac{\Delta_0}{h}}^{\nu_{max}}\rho(\nu)n(\nu,T)d\nu \ ,
    \label{POPOL}
\end{equation}
where $\bar{n}$ is the average phonon population, $\rho(\nu)$ is the phonon DOS, and $n(\nu,T)$ is the Bose-Einstein population for fixed frequency and temperature.
The infrared cutoff is given by the minimum energy necessary to break the cooper pairs, corresponding to the superconductor gap $\Delta_0$, while the other one is given by the maximum phonon frequency of the material ($\nu_{max}$). 
Taxing as an example aluminium as the superconductor ($\Delta_0\approx\SI{170}{\micro\electronvolt}$~\cite{KITTEL}) and a cubic CsPbI$_3$ crystal of \SI{1}{\centi\meter} edge (about 10\textsuperscript{21} crystal cells) we calculated the average thermal population above the gap. 
At a temperature of \SI{10}{\milli\kelvin}, achievable with a dilution cryostat, the average population is about 10\textsuperscript{-70} phonons above the energy gap. By estimating a mean lifetime of the optical phonons of \SI{1}{\pico\second}, the expected number of events in one year is about $10^{-50}$, corresponding to a completely negligible thermal background for the single phonon revelation.
Moreover, the breaking of the cooper pairs can be achieved also by the acoustic phonons generated by the decay of the optical one, increasing the interaction time with the superconducting resonator to the ns-\si{\micro\second} range.

\section{Conclusions}\label{sec:sec5}
In this work, we have explored the potential of metal halide perovskites as promising targets for DM direct detection probes via single-phonon excitations, both in the case of a dark photon mediated scattering and absorption. These materials, with their unique vibrational properties and anisotropic lattice structures, exhibit several gapped optical phonons with energies in the range of few meV. Our results put the inorganic CsPbI$_3$ and hybrids MaPbI$_3$ and MaPbBr$_3$ into the stage as their sensitivities are comparable to, if not better than, other polar materials already studied in the literature, such as CsI, GaAs, and the anisotropic Al$_2$O$_3$. 

\smallskip
Regarding the dark photon mediated scattering process, the main results are summarized in Sec.~\ref{sec:sec2}. We found that the Fr\"ohlich narrow width approximation, provides scattering rates compatible with the ab-initio DFT approach. In particular metal halide perovskites achieve a sensitivity that is improved by two orders of magnitude with respect to Al$_2$O$_3$ and GaAs, with performance comparable to CsI. Specifically, the peak performance is achieved with \(\mathrm{MaPbI}_3\), reaching a sensitivity of $\Bar{\sigma}_e=4\cdot 10^{-45}\,\si{cm^2}$ at $m_\chi=14$ keV. The lower phonon gap of these materials enables new kinematic possibilities, allowing for probes with Earth-based future experiments of the keV mass scale, related to a warm DM candidate.

Fig.~\ref{fig:ratemodes} shows the interplay of the phonon modes that contribute to the scattering cross section. In particular we focus more on the inorganic CsPbI$_3$ as it is expected to suffer less from degradation compared to hybrid perovskites. Due to its anisotropic structure, CsPbI$_3$ also facilitates daily modulation analysis, which is not possible in isotropic material like CsI. We found that CsPbI$_3$ exhibits a significant modulation for DM masses around $\approx1$ keV, whereas modulation becomes statistically insignificant for masses $\gtrsim 10$ keV. However, complementary results are obtained by considering a different crystal structure, $\delta$-CsPbI$_3$, which allows probing the modulating signal up to 40 keV. In conclusion, CsPbI$_3$ stands out as the most effective material in terms of total detection rate, with its daily modulation signal offering competitive results compared to other materials analyzed in the literature.

\smallskip
In Sec.~\ref{sec:sec3}, we present the results for the dark photon absorption process. The absorption rate per unit target was calculated using the ELF in the zero-momentum limit. In Fig.~\ref{fig:Abs}, we show the 95\% CL sensitivity on the \(\kappa\) mixing parameter as a function of the dark photon mass \(m_{A'}\). Our findings indicate that metal halide perovskites demonstrate exceptional sensitivity compared to other targets, setting the most stringent limits for dark photon masses below 8 meV.

\smallskip
In Sec.~\ref{sec:sec4} we propose a possible experimental setup. We found that it is possible to grow, as already demonstrated in the literature, kilogram-scale single crystals of CsPbI$_3$. This scalability, combined with the intrinsic stability of CsPbI$_3$, makes it an excellent candidate for the development of detectors targeting light DM through single-phonon emission. Moreover, the operation at temperature around \SI{10}{\milli\kelvin} in a dilution cryostat can heavily suppress the background from thermal phonons, allowing in principle the revelation of the  signal from a single DM particle interaction.

\smallskip
Our findings highlight the remarkable potential of metal halide perovskites in advancing direct DM searches, particularly for exploring light DM masses in an uncharted region of parameter space accessible to terrestrial experiments. Their unique phononic properties, high sensitivity, stability, and ease of construction offer distinct advantages, representing a significant step forward in the search for light DM.

\section{Acknowledgements}
We would like to thank Angelo Esposito for insightful discussions and a thorough review of the manuscript, and Francesco Pineider for fruitful discussions. We thanks Andrea Mitridate and Tanner Trickle for useful discussions. This work is partially supported by the U.S. Department of Energy, Office of Science, National Quantum Information Science Research Centers, Superconducting Quantum Materials and Systems Center (SQMS) under contract number DE-AC02-07CH11359. The research conducted by Giulio Marino and Paolo Panci
receives partial funding from the European Union–Next
generation EU (through Progetti di Ricerca di Interesse
Nazionale (PRIN) Grant No. 202289JEW4). 

\appendix
\section{$\delta$-CsPbI$_3$}
\label{app:app1}

\begin{figure}[h!!!]
 \centering
    \includegraphics[width=0.7\columnwidth]{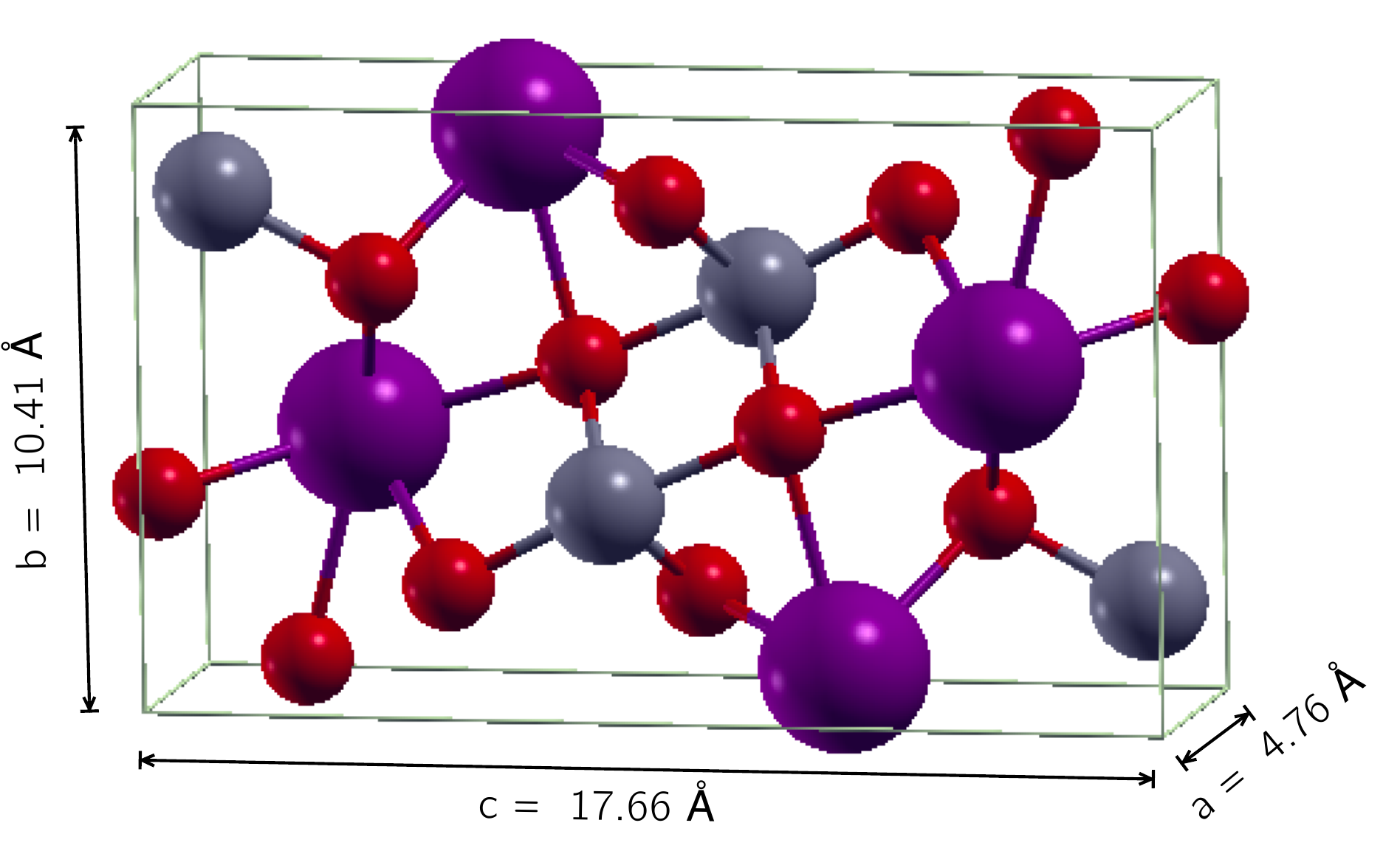}
    \vspace{0.cm}
    \centering
    \includegraphics[width=0.67\columnwidth]{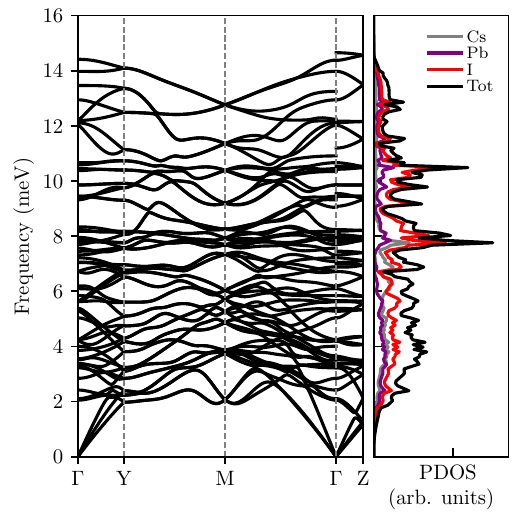}
    \caption{\textbf{Top panel:} Crystal structure of \(\delta\)-CsPbI\(_3\) with lattice parameters, shown from both top and lateral views. \textbf{Bottom panel:} Phonon dispersion on the $\Gamma$-Y-M-$\Gamma$-Z high symmetry path (left) and PDOS (right).}
    \label{fig:delta-cspbi3}
\end{figure}

Hereafter, we provide the details of the ab-initio calculations for the \(\delta\)-CsPbI\(_3\) phase. As with the \(\gamma\)-CsPbI\(_3\) phase described in Sec.~\ref{sec:DFTanalysis}, all calculations are performed using the Quantum Espresso suite \cite{QE-2009,QE-2017}. Projector-augmented wave pseudopotentials are employed with the PBEsol exchange-correlation functional, using cutoffs of 65 Ry and 360 Ry for the wave functions and the charge density, respectively.
We perform structural relaxation until the forces for all atoms are smaller than \(10^{-4} \, \text{Ry}/\text{\AA}\), sampling the Brillouin zone with an 11x5x3 \(\Gamma\)-centered MP grid \cite{MPgrid}. The equilibrium lattice constants are determined as \(a = 4.76 \, \text{\AA}\), \(b = 10.41 \, \text{\AA}\), and \(c = 17.66 \, \text{\AA}\), in good agreement with previous reports \cite{GammaCsPbI3_Lattice2_MatPhyToday_2023}, as shown in the top panel of Fig.~\ref{fig:delta-cspbi3}. Self-consistent calculations are performed on the same grid with a convergence threshold of \(10^{-12} \, \text{Ry}\).  
Phonon dispersion is obtained using the frozen phonon method as implemented in \texttt{Phonopy} \cite{phonopy-phono3py-JPCM,phonopy-phono3py-JPSJ}, with a 3x2x1 supercell and a 2x2x2 \(\Gamma\)-centered MP grid, including non-analytical term corrections \cite{nac1,nac2,nac3}. The phonon dispersion relation and the Projected Density of States (PDOS) for each atomic species are shown in the bottom panel of Fig.~\ref{fig:delta-cspbi3}.

\newpage

\bigskip
 \bibliography{biblio.bib}

\begin{thebibliography}{70}%
\makeatletter
\providecommand \@ifxundefined [1]{%
 \@ifx{#1\undefined}
}%
\providecommand \@ifnum [1]{%
 \ifnum #1\expandafter \@firstoftwo
 \else \expandafter \@secondoftwo
 \fi
}%
\providecommand \@ifx [1]{%
 \ifx #1\expandafter \@firstoftwo
 \else \expandafter \@secondoftwo
 \fi
}%
\providecommand \natexlab [1]{#1}%
\providecommand \enquote  [1]{``#1''}%
\providecommand \bibnamefont  [1]{#1}%
\providecommand \bibfnamefont [1]{#1}%
\providecommand \citenamefont [1]{#1}%
\providecommand \href@noop [0]{\@secondoftwo}%
\providecommand \href [0]{\begingroup \@sanitize@url \@href}%
\providecommand \@href[1]{\@@startlink{#1}\@@href}%
\providecommand \@@href[1]{\endgroup#1\@@endlink}%
\providecommand \@sanitize@url [0]{\catcode `\\12\catcode `\$12\catcode
  `\&12\catcode `\#12\catcode `\^12\catcode `\_12\catcode `\%12\relax}%
\providecommand \@@startlink[1]{}%
\providecommand \@@endlink[0]{}%
\providecommand \url  [0]{\begingroup\@sanitize@url \@url }%
\providecommand \@url [1]{\endgroup\@href {#1}{\urlprefix }}%
\providecommand \urlprefix  [0]{URL }%
\providecommand \Eprint [0]{\href }%
\providecommand \doibase [0]{http://dx.doi.org/}%
\providecommand \selectlanguage [0]{\@gobble}%
\providecommand \bibinfo  [0]{\@secondoftwo}%
\providecommand \bibfield  [0]{\@secondoftwo}%
\providecommand \translation [1]{[#1]}%
\providecommand \BibitemOpen [0]{}%
\providecommand \bibitemStop [0]{}%
\providecommand \bibitemNoStop [0]{.\EOS\space}%
\providecommand \EOS [0]{\spacefactor3000\relax}%
\providecommand \BibitemShut  [1]{\csname bibitem#1\endcsname}%
\let\auto@bib@innerbib\@empty
\bibitem [{\citenamefont {Aprile}\ and\ \citenamefont {the
  XENON~Collaboration}(2019)}]{PhysRevLett.123.251801}%
  \BibitemOpen
  \bibfield  {author} {\bibinfo {author} {\bibfnamefont {E.}~\bibnamefont
  {Aprile}}\ and\ \bibinfo {author} {\bibnamefont {the XENON~Collaboration}},\
  }\href {\doibase https://doi.org/10.1103/PhysRevLett.123.251801} {\bibfield
  {journal} {\bibinfo  {journal} {Phys. Rev. Lett.}\ }\textbf {\bibinfo
  {volume} {123}},\ \bibinfo {pages} {251801} (\bibinfo {year}
  {2019})}\BibitemShut {NoStop}%
\bibitem [{\citenamefont {Hall}\ \emph {et~al.}(2010)\citenamefont {Hall},
  \citenamefont {Jedamzik}, \citenamefont {March-Russell},\ and\ \citenamefont
  {West}}]{Hall_2010}%
  \BibitemOpen
  \bibfield  {author} {\bibinfo {author} {\bibfnamefont {L.~J.}\ \bibnamefont
  {Hall}}, \bibinfo {author} {\bibfnamefont {K.}~\bibnamefont {Jedamzik}},
  \bibinfo {author} {\bibfnamefont {J.}~\bibnamefont {March-Russell}}, \ and\
  \bibinfo {author} {\bibfnamefont {S.~M.}\ \bibnamefont {West}},\ }\href
  {\doibase 10.1007/jhep03(2010)080} {\bibfield  {journal} {\bibinfo  {journal}
  {Journal of High Energy Physics}\ }\textbf {\bibinfo {volume} {2010}}
  (\bibinfo {year} {2010}),\ 10.1007/jhep03(2010)080}\BibitemShut {NoStop}%
\bibitem [{\citenamefont {Bernal}\ \emph {et~al.}(2017)\citenamefont {Bernal},
  \citenamefont {Heikinheimo}, \citenamefont {Tenkanen}, \citenamefont
  {Tuominen},\ and\ \citenamefont {Vaskonen}}]{Bernal_2017}%
  \BibitemOpen
  \bibfield  {author} {\bibinfo {author} {\bibfnamefont {N.}~\bibnamefont
  {Bernal}}, \bibinfo {author} {\bibfnamefont {M.}~\bibnamefont {Heikinheimo}},
  \bibinfo {author} {\bibfnamefont {T.}~\bibnamefont {Tenkanen}}, \bibinfo
  {author} {\bibfnamefont {K.}~\bibnamefont {Tuominen}}, \ and\ \bibinfo
  {author} {\bibfnamefont {V.}~\bibnamefont {Vaskonen}},\ }\href {\doibase
  https://doi.org/10.1142/S0217751X1730023X} {\bibfield  {journal} {\bibinfo
  {journal} {International Journal of Modern Physics A}\ }\textbf {\bibinfo
  {volume} {32}},\ \bibinfo {pages} {1730023} (\bibinfo {year}
  {2017})}\BibitemShut {NoStop}%
\bibitem [{\citenamefont {Hooper}\ and\ \citenamefont {Zurek}(2008)}]{hidden1}%
  \BibitemOpen
  \bibfield  {author} {\bibinfo {author} {\bibfnamefont {D.}~\bibnamefont
  {Hooper}}\ and\ \bibinfo {author} {\bibfnamefont {K.~M.}\ \bibnamefont
  {Zurek}},\ }\href {\doibase https://doi.org/10.1103/PhysRevD.77.087302}
  {\bibfield  {journal} {\bibinfo  {journal} {Phys. Rev. D}\ }\textbf {\bibinfo
  {volume} {77}},\ \bibinfo {pages} {087302} (\bibinfo {year}
  {2008})}\BibitemShut {NoStop}%
\bibitem [{\citenamefont {Feng}\ and\ \citenamefont {Kumar}(2008)}]{hidden2}%
  \BibitemOpen
  \bibfield  {author} {\bibinfo {author} {\bibfnamefont {J.~L.}\ \bibnamefont
  {Feng}}\ and\ \bibinfo {author} {\bibfnamefont {J.}~\bibnamefont {Kumar}},\
  }\href {\doibase https://doi.org/10.1103/PhysRevLett.101.231301} {\bibfield
  {journal} {\bibinfo  {journal} {Phys. Rev. Lett.}\ }\textbf {\bibinfo
  {volume} {101}},\ \bibinfo {pages} {231301} (\bibinfo {year}
  {2008})}\BibitemShut {NoStop}%
\bibitem [{\citenamefont {Cohen}\ \emph {et~al.}(2010)\citenamefont {Cohen},
  \citenamefont {Phalen}, \citenamefont {Pierce},\ and\ \citenamefont
  {Zurek}}]{hidden3}%
  \BibitemOpen
  \bibfield  {author} {\bibinfo {author} {\bibfnamefont {T.}~\bibnamefont
  {Cohen}}, \bibinfo {author} {\bibfnamefont {D.~J.}\ \bibnamefont {Phalen}},
  \bibinfo {author} {\bibfnamefont {A.}~\bibnamefont {Pierce}}, \ and\ \bibinfo
  {author} {\bibfnamefont {K.~M.}\ \bibnamefont {Zurek}},\ }\href {\doibase
  https://doi.org/10.1103/PhysRevD.82.056001} {\bibfield  {journal} {\bibinfo
  {journal} {Phys. Rev. D}\ }\textbf {\bibinfo {volume} {82}},\ \bibinfo
  {pages} {056001} (\bibinfo {year} {2010})}\BibitemShut {NoStop}%
\bibitem [{\citenamefont {Schutz}\ and\ \citenamefont
  {Zurek}(2016)}]{Singleph1}%
  \BibitemOpen
  \bibfield  {author} {\bibinfo {author} {\bibfnamefont {K.}~\bibnamefont
  {Schutz}}\ and\ \bibinfo {author} {\bibfnamefont {K.~M.}\ \bibnamefont
  {Zurek}},\ }\href {\doibase https://doi.org/10.1103/PhysRevLett.117.121302}
  {\bibfield  {journal} {\bibinfo  {journal} {Phys. Rev. Lett.}\ }\textbf
  {\bibinfo {volume} {117}},\ \bibinfo {pages} {121302} (\bibinfo {year}
  {2016})}\BibitemShut {NoStop}%
\bibitem [{\citenamefont {Knapen}\ \emph
  {et~al.}(2017{\natexlab{a}})\citenamefont {Knapen}, \citenamefont {Lin},\
  and\ \citenamefont {Zurek}}]{Singleph2}%
  \BibitemOpen
  \bibfield  {author} {\bibinfo {author} {\bibfnamefont {S.}~\bibnamefont
  {Knapen}}, \bibinfo {author} {\bibfnamefont {T.}~\bibnamefont {Lin}}, \ and\
  \bibinfo {author} {\bibfnamefont {K.~M.}\ \bibnamefont {Zurek}},\ }\href
  {\doibase https://doi.org/10.1103/PhysRevD.95.056019} {\bibfield  {journal}
  {\bibinfo  {journal} {Phys. Rev. D}\ }\textbf {\bibinfo {volume} {95}},\
  \bibinfo {pages} {056019} (\bibinfo {year} {2017}{\natexlab{a}})}\BibitemShut
  {NoStop}%
\bibitem [{\citenamefont {Caputo}\ \emph {et~al.}(2019)\citenamefont {Caputo},
  \citenamefont {Esposito},\ and\ \citenamefont {Polosa}}]{Singleph3}%
  \BibitemOpen
  \bibfield  {author} {\bibinfo {author} {\bibfnamefont {A.}~\bibnamefont
  {Caputo}}, \bibinfo {author} {\bibfnamefont {A.}~\bibnamefont {Esposito}}, \
  and\ \bibinfo {author} {\bibfnamefont {A.~D.}\ \bibnamefont {Polosa}},\
  }\href {\doibase https://doi.org/10.1103/PhysRevD.100.116007} {\bibfield
  {journal} {\bibinfo  {journal} {Phys. Rev. D}\ }\textbf {\bibinfo {volume}
  {100}},\ \bibinfo {pages} {116007} (\bibinfo {year} {2019})}\BibitemShut
  {NoStop}%
\bibitem [{\citenamefont {Baym}\ \emph {et~al.}(2020)\citenamefont {Baym},
  \citenamefont {Beck}, \citenamefont {Filippini}, \citenamefont {Pethick},\
  and\ \citenamefont {Shelton}}]{Singleph4}%
  \BibitemOpen
  \bibfield  {author} {\bibinfo {author} {\bibfnamefont {G.}~\bibnamefont
  {Baym}}, \bibinfo {author} {\bibfnamefont {D.~H.}\ \bibnamefont {Beck}},
  \bibinfo {author} {\bibfnamefont {J.~P.}\ \bibnamefont {Filippini}}, \bibinfo
  {author} {\bibfnamefont {C.~J.}\ \bibnamefont {Pethick}}, \ and\ \bibinfo
  {author} {\bibfnamefont {J.}~\bibnamefont {Shelton}},\ }\href {\doibase
  https://doi.org/10.1103/PhysRevD.102.035014} {\bibfield  {journal} {\bibinfo
  {journal} {Phys. Rev. D}\ }\textbf {\bibinfo {volume} {102}},\ \bibinfo
  {pages} {035014} (\bibinfo {year} {2020})}\BibitemShut {NoStop}%
\bibitem [{\citenamefont {Caputo}\ \emph {et~al.}(2021)\citenamefont {Caputo},
  \citenamefont {Esposito}, \citenamefont {Piccinini}, \citenamefont {Polosa},\
  and\ \citenamefont {Rossi}}]{Singleph5}%
  \BibitemOpen
  \bibfield  {author} {\bibinfo {author} {\bibfnamefont {A.}~\bibnamefont
  {Caputo}}, \bibinfo {author} {\bibfnamefont {A.}~\bibnamefont {Esposito}},
  \bibinfo {author} {\bibfnamefont {F.}~\bibnamefont {Piccinini}}, \bibinfo
  {author} {\bibfnamefont {A.~D.}\ \bibnamefont {Polosa}}, \ and\ \bibinfo
  {author} {\bibfnamefont {G.}~\bibnamefont {Rossi}},\ }\href {\doibase
  https://doi.org/10.1103/PhysRevD.103.055017} {\bibfield  {journal} {\bibinfo
  {journal} {Phys. Rev. D}\ }\textbf {\bibinfo {volume} {103}},\ \bibinfo
  {pages} {055017} (\bibinfo {year} {2021})}\BibitemShut {NoStop}%
\bibitem [{\citenamefont {Griffin}\ \emph {et~al.}(2018)\citenamefont
  {Griffin}, \citenamefont {Knapen}, \citenamefont {Lin},\ and\ \citenamefont
  {Zurek}}]{ZurekPRD2018}%
  \BibitemOpen
  \bibfield  {author} {\bibinfo {author} {\bibfnamefont {S.}~\bibnamefont
  {Griffin}}, \bibinfo {author} {\bibfnamefont {S.}~\bibnamefont {Knapen}},
  \bibinfo {author} {\bibfnamefont {T.}~\bibnamefont {Lin}}, \ and\ \bibinfo
  {author} {\bibfnamefont {K.~M.}\ \bibnamefont {Zurek}},\ }\href {\doibase
  10.1103/PhysRevD.98.115034} {\bibfield  {journal} {\bibinfo  {journal} {Phys.
  Rev. D}\ }\textbf {\bibinfo {volume} {98}},\ \bibinfo {pages} {115034}
  (\bibinfo {year} {2018})}\BibitemShut {NoStop}%
\bibitem [{\citenamefont {Griffin}\ \emph {et~al.}(2020)\citenamefont
  {Griffin}, \citenamefont {Inzani}, \citenamefont {Trickle}, \citenamefont
  {Zhang},\ and\ \citenamefont {Zurek}}]{epsELF2}%
  \BibitemOpen
  \bibfield  {author} {\bibinfo {author} {\bibfnamefont {S.~M.}\ \bibnamefont
  {Griffin}}, \bibinfo {author} {\bibfnamefont {K.}~\bibnamefont {Inzani}},
  \bibinfo {author} {\bibfnamefont {T.}~\bibnamefont {Trickle}}, \bibinfo
  {author} {\bibfnamefont {Z.}~\bibnamefont {Zhang}}, \ and\ \bibinfo {author}
  {\bibfnamefont {K.~M.}\ \bibnamefont {Zurek}},\ }\href {\doibase
  https://doi.org/10.1103/PhysRevD.101.055004} {\bibfield  {journal} {\bibinfo
  {journal} {Phys. Rev. D}\ }\textbf {\bibinfo {volume} {101}},\ \bibinfo
  {pages} {055004} (\bibinfo {year} {2020})}\BibitemShut {NoStop}%
\bibitem [{\citenamefont {Campbell-Deem}\ \emph {et~al.}(2020)\citenamefont
  {Campbell-Deem}, \citenamefont {Cox}, \citenamefont {Knapen}, \citenamefont
  {Lin},\ and\ \citenamefont {Melia}}]{Singleph8}%
  \BibitemOpen
  \bibfield  {author} {\bibinfo {author} {\bibfnamefont {B.}~\bibnamefont
  {Campbell-Deem}}, \bibinfo {author} {\bibfnamefont {P.}~\bibnamefont {Cox}},
  \bibinfo {author} {\bibfnamefont {S.}~\bibnamefont {Knapen}}, \bibinfo
  {author} {\bibfnamefont {T.}~\bibnamefont {Lin}}, \ and\ \bibinfo {author}
  {\bibfnamefont {T.}~\bibnamefont {Melia}},\ }\href {\doibase
  https://doi.org/10.1103/PhysRevD.102.019904} {\bibfield  {journal} {\bibinfo
  {journal} {Phys. Rev. D}\ }\textbf {\bibinfo {volume} {102}},\ \bibinfo
  {pages} {019904} (\bibinfo {year} {2020})}\BibitemShut {NoStop}%
\bibitem [{\citenamefont {Griffin}\ \emph {et~al.}(2021)\citenamefont
  {Griffin}, \citenamefont {Hochberg}, \citenamefont {Inzani}, \citenamefont
  {Kurinsky}, \citenamefont {Lin},\ and\ \citenamefont {Yu}}]{Singleph9}%
  \BibitemOpen
  \bibfield  {author} {\bibinfo {author} {\bibfnamefont {S.~M.}\ \bibnamefont
  {Griffin}}, \bibinfo {author} {\bibfnamefont {Y.}~\bibnamefont {Hochberg}},
  \bibinfo {author} {\bibfnamefont {K.}~\bibnamefont {Inzani}}, \bibinfo
  {author} {\bibfnamefont {N.}~\bibnamefont {Kurinsky}}, \bibinfo {author}
  {\bibfnamefont {T.}~\bibnamefont {Lin}}, \ and\ \bibinfo {author}
  {\bibfnamefont {T.~C.}\ \bibnamefont {Yu}},\ }\href {\doibase
  https://doi.org/10.1103/PhysRevD.103.075002} {\bibfield  {journal} {\bibinfo
  {journal} {Phys. Rev. D}\ }\textbf {\bibinfo {volume} {103}},\ \bibinfo
  {pages} {075002} (\bibinfo {year} {2021})}\BibitemShut {NoStop}%
\bibitem [{\citenamefont {Trickle}\ \emph {et~al.}(2021)\citenamefont
  {Trickle}, \citenamefont {Zhang},\ and\ \citenamefont {Zurek}}]{Singleph10}%
  \BibitemOpen
  \bibfield  {author} {\bibinfo {author} {\bibfnamefont {T.}~\bibnamefont
  {Trickle}}, \bibinfo {author} {\bibfnamefont {Z.}~\bibnamefont {Zhang}}, \
  and\ \bibinfo {author} {\bibfnamefont {K.~M.}\ \bibnamefont {Zurek}},\ }\href
  {https://arxiv.org/abs/2009.13534} {\enquote {\bibinfo {title} {Effective
  field theory of dark matter direct detection with collective excitations},}\
  } (\bibinfo {year} {2021}),\ \Eprint {http://arxiv.org/abs/2009.13534}
  {arXiv:2009.13534 [hep-ph]} \BibitemShut {NoStop}%
\bibitem [{\citenamefont {Kahn}\ \emph {et~al.}(2021)\citenamefont {Kahn},
  \citenamefont {Krnjaic},\ and\ \citenamefont {Mandava}}]{Singleph11}%
  \BibitemOpen
  \bibfield  {author} {\bibinfo {author} {\bibfnamefont {Y.}~\bibnamefont
  {Kahn}}, \bibinfo {author} {\bibfnamefont {G.}~\bibnamefont {Krnjaic}}, \
  and\ \bibinfo {author} {\bibfnamefont {B.}~\bibnamefont {Mandava}},\ }\href
  {\doibase https://doi.org/10.1103/PhysRevLett.127.081804} {\bibfield
  {journal} {\bibinfo  {journal} {Phys. Rev. Lett.}\ }\textbf {\bibinfo
  {volume} {127}},\ \bibinfo {pages} {081804} (\bibinfo {year}
  {2021})}\BibitemShut {NoStop}%
\bibitem [{\citenamefont {Knapen}\ \emph {et~al.}(2021)\citenamefont {Knapen},
  \citenamefont {Kozaczuk},\ and\ \citenamefont {Lin}}]{Singleph12}%
  \BibitemOpen
  \bibfield  {author} {\bibinfo {author} {\bibfnamefont {S.}~\bibnamefont
  {Knapen}}, \bibinfo {author} {\bibfnamefont {J.}~\bibnamefont {Kozaczuk}}, \
  and\ \bibinfo {author} {\bibfnamefont {T.}~\bibnamefont {Lin}},\ }\href
  {\doibase https://doi.org/10.1103/PhysRevLett.127.081805} {\bibfield
  {journal} {\bibinfo  {journal} {Phys. Rev. Lett.}\ }\textbf {\bibinfo
  {volume} {127}},\ \bibinfo {pages} {081805} (\bibinfo {year}
  {2021})}\BibitemShut {NoStop}%
\bibitem [{\citenamefont {Acanfora}\ \emph {et~al.}(2019)\citenamefont
  {Acanfora}, \citenamefont {Esposito},\ and\ \citenamefont
  {Polosa}}]{Singleph13}%
  \BibitemOpen
  \bibfield  {author} {\bibinfo {author} {\bibfnamefont {F.}~\bibnamefont
  {Acanfora}}, \bibinfo {author} {\bibfnamefont {A.}~\bibnamefont {Esposito}},
  \ and\ \bibinfo {author} {\bibfnamefont {A.~D.}\ \bibnamefont {Polosa}},\
  }\href {\doibase 10.1140/epjc/s10052-019-7057-0} {\bibfield  {journal}
  {\bibinfo  {journal} {The European Physical Journal C}\ }\textbf {\bibinfo
  {volume} {79}},\ \bibinfo {pages} {549} (\bibinfo {year} {2019})}\BibitemShut
  {NoStop}%
\bibitem [{\citenamefont {Trickle}\ \emph {et~al.}(2020)\citenamefont
  {Trickle}, \citenamefont {Zhang}, \citenamefont {Zurek}, \citenamefont
  {Inzani},\ and\ \citenamefont {Griffin}}]{Trickle_2020}%
  \BibitemOpen
  \bibfield  {author} {\bibinfo {author} {\bibfnamefont {T.}~\bibnamefont
  {Trickle}}, \bibinfo {author} {\bibfnamefont {Z.}~\bibnamefont {Zhang}},
  \bibinfo {author} {\bibfnamefont {K.~M.}\ \bibnamefont {Zurek}}, \bibinfo
  {author} {\bibfnamefont {K.}~\bibnamefont {Inzani}}, \ and\ \bibinfo {author}
  {\bibfnamefont {S.~M.}\ \bibnamefont {Griffin}},\ }\href {\doibase
  https://doi.org/10.1007/JHEP03(2020)036} {\bibfield  {journal} {\bibinfo
  {journal} {Journal of High Energy Physics}\ }\textbf {\bibinfo {volume}
  {2020}},\ \bibinfo {pages} {36} (\bibinfo {year} {2020})}\BibitemShut
  {NoStop}%
\bibitem [{\citenamefont {Paolucci}\ and\ \citenamefont
  {Giazotto}(2021)}]{PAOLUCCISUPERCONDUCTING}%
  \BibitemOpen
  \bibfield  {author} {\bibinfo {author} {\bibfnamefont {F.}~\bibnamefont
  {Paolucci}}\ and\ \bibinfo {author} {\bibfnamefont {F.}~\bibnamefont
  {Giazotto}},\ }\href {\doibase 10.3390/instruments5020014} {\bibfield
  {journal} {\bibinfo  {journal} {Instruments}\ }\textbf {\bibinfo {volume}
  {5}} (\bibinfo {year} {2021}),\ 10.3390/instruments5020014}\BibitemShut
  {NoStop}%
\bibitem [{\citenamefont {Paolucci}\ \emph {et~al.}(2020)\citenamefont
  {Paolucci}, \citenamefont {Buccheri}, \citenamefont {Germanese},
  \citenamefont {Ligato}, \citenamefont {Paoletti}, \citenamefont {Signorelli},
  \citenamefont {Bitossi}, \citenamefont {Spagnolo}, \citenamefont {Falferi},
  \citenamefont {Rajteri}, \citenamefont {Gatti},\ and\ \citenamefont
  {Giazotto}}]{PaolucciTES_DM_2020}%
  \BibitemOpen
  \bibfield  {author} {\bibinfo {author} {\bibfnamefont {F.}~\bibnamefont
  {Paolucci}}, \bibinfo {author} {\bibfnamefont {V.}~\bibnamefont {Buccheri}},
  \bibinfo {author} {\bibfnamefont {G.}~\bibnamefont {Germanese}}, \bibinfo
  {author} {\bibfnamefont {N.}~\bibnamefont {Ligato}}, \bibinfo {author}
  {\bibfnamefont {R.}~\bibnamefont {Paoletti}}, \bibinfo {author}
  {\bibfnamefont {G.}~\bibnamefont {Signorelli}}, \bibinfo {author}
  {\bibfnamefont {M.}~\bibnamefont {Bitossi}}, \bibinfo {author} {\bibfnamefont
  {P.}~\bibnamefont {Spagnolo}}, \bibinfo {author} {\bibfnamefont
  {P.}~\bibnamefont {Falferi}}, \bibinfo {author} {\bibfnamefont
  {M.}~\bibnamefont {Rajteri}}, \bibinfo {author} {\bibfnamefont
  {C.}~\bibnamefont {Gatti}}, \ and\ \bibinfo {author} {\bibfnamefont
  {F.}~\bibnamefont {Giazotto}},\ }\href {\doibase 10.1063/5.0021996}
  {\bibfield  {journal} {\bibinfo  {journal} {Journal of Applied Physics}\
  }\textbf {\bibinfo {volume} {128}},\ \bibinfo {pages} {194502} (\bibinfo
  {year} {2020})},\ \Eprint
  {http://arxiv.org/abs/https://pubs.aip.org/aip/jap/article-pdf/doi/10.1063/5.0021996/15255794/194502\_1\_online.pdf}
  {https://pubs.aip.org/aip/jap/article-pdf/doi/10.1063/5.0021996/15255794/194502\_1\_online.pdf}
  \BibitemShut {NoStop}%
\bibitem [{\citenamefont {Geilhufe}\ \emph {et~al.}(2020)\citenamefont
  {Geilhufe}, \citenamefont {Kahlhoefer},\ and\ \citenamefont
  {Winkler}}]{DiracMat_PhysRevD.101.055005}%
  \BibitemOpen
  \bibfield  {author} {\bibinfo {author} {\bibfnamefont {R.~M.}\ \bibnamefont
  {Geilhufe}}, \bibinfo {author} {\bibfnamefont {F.}~\bibnamefont
  {Kahlhoefer}}, \ and\ \bibinfo {author} {\bibfnamefont {M.~W.}\ \bibnamefont
  {Winkler}},\ }\href {\doibase 10.1103/PhysRevD.101.055005} {\bibfield
  {journal} {\bibinfo  {journal} {Phys. Rev. D}\ }\textbf {\bibinfo {volume}
  {101}},\ \bibinfo {pages} {055005} (\bibinfo {year} {2020})}\BibitemShut
  {NoStop}%
\bibitem [{\citenamefont {Coskuner}\ \emph {et~al.}(2021)\citenamefont
  {Coskuner}, \citenamefont {Mitridate}, \citenamefont {Olivares},\ and\
  \citenamefont {Zurek}}]{DirectionalPRDZurek}%
  \BibitemOpen
  \bibfield  {author} {\bibinfo {author} {\bibfnamefont {A.}~\bibnamefont
  {Coskuner}}, \bibinfo {author} {\bibfnamefont {A.}~\bibnamefont {Mitridate}},
  \bibinfo {author} {\bibfnamefont {A.}~\bibnamefont {Olivares}}, \ and\
  \bibinfo {author} {\bibfnamefont {K.~M.}\ \bibnamefont {Zurek}},\ }\href
  {\doibase https://doi.org/10.1103/PhysRevD.103.016006} {\bibfield  {journal}
  {\bibinfo  {journal} {Phys. Rev. D}\ }\textbf {\bibinfo {volume} {103}},\
  \bibinfo {pages} {016006} (\bibinfo {year} {2021})}\BibitemShut {NoStop}%
\bibitem [{\citenamefont {Coskuner}\ \emph {et~al.}(2022)\citenamefont
  {Coskuner}, \citenamefont {Trickle}, \citenamefont {Zhang},\ and\
  \citenamefont {Zurek}}]{TargCompSinglph}%
  \BibitemOpen
  \bibfield  {author} {\bibinfo {author} {\bibfnamefont {A.}~\bibnamefont
  {Coskuner}}, \bibinfo {author} {\bibfnamefont {T.}~\bibnamefont {Trickle}},
  \bibinfo {author} {\bibfnamefont {Z.}~\bibnamefont {Zhang}}, \ and\ \bibinfo
  {author} {\bibfnamefont {K.~M.}\ \bibnamefont {Zurek}},\ }\href {\doibase
  https://doi.org/10.1103/PhysRevD.105.015010} {\bibfield  {journal} {\bibinfo
  {journal} {Phys. Rev. D}\ }\textbf {\bibinfo {volume} {105}},\ \bibinfo
  {pages} {015010} (\bibinfo {year} {2022})}\BibitemShut {NoStop}%
\bibitem [{\citenamefont {Herz}(2017)}]{cite-key}%
  \BibitemOpen
  \bibfield  {author} {\bibinfo {author} {\bibfnamefont {L.~M.}\ \bibnamefont
  {Herz}},\ }\href {\doibase 10.1021/acsenergylett.7b00276} {\bibfield
  {journal} {\bibinfo  {journal} {ACS Energy Letters}\ }\textbf {\bibinfo
  {volume} {2}},\ \bibinfo {pages} {1539} (\bibinfo {year} {2017})}\BibitemShut
  {NoStop}%
\bibitem [{\citenamefont {Freese}\ \emph {et~al.}(1988)\citenamefont {Freese},
  \citenamefont {Frieman},\ and\ \citenamefont {Gould}}]{dailyMod1}%
  \BibitemOpen
  \bibfield  {author} {\bibinfo {author} {\bibfnamefont {K.}~\bibnamefont
  {Freese}}, \bibinfo {author} {\bibfnamefont {J.}~\bibnamefont {Frieman}}, \
  and\ \bibinfo {author} {\bibfnamefont {A.}~\bibnamefont {Gould}},\ }\href
  {\doibase 10.1103/PhysRevD.37.3388} {\bibfield  {journal} {\bibinfo
  {journal} {Phys. Rev. D}\ }\textbf {\bibinfo {volume} {37}},\ \bibinfo
  {pages} {3388} (\bibinfo {year} {1988})}\BibitemShut {NoStop}%
\bibitem [{\citenamefont {Drukier}\ \emph {et~al.}(1986)\citenamefont
  {Drukier}, \citenamefont {Freese},\ and\ \citenamefont
  {Spergel}}]{dailyMod2}%
  \BibitemOpen
  \bibfield  {author} {\bibinfo {author} {\bibfnamefont {A.~K.}\ \bibnamefont
  {Drukier}}, \bibinfo {author} {\bibfnamefont {K.}~\bibnamefont {Freese}}, \
  and\ \bibinfo {author} {\bibfnamefont {D.~N.}\ \bibnamefont {Spergel}},\
  }\href {\doibase 10.1103/PhysRevD.33.3495} {\bibfield  {journal} {\bibinfo
  {journal} {Phys. Rev. D}\ }\textbf {\bibinfo {volume} {33}},\ \bibinfo
  {pages} {3495} (\bibinfo {year} {1986})}\BibitemShut {NoStop}%
\bibitem [{\citenamefont {Toloueinia}\ \emph {et~al.}(2020)\citenamefont
  {Toloueinia}, \citenamefont {Khassaf}, \citenamefont {Shirazi~Amin},
  \citenamefont {Tobin}, \citenamefont {Alpay},\ and\ \citenamefont
  {Suib}}]{MAPIdegradation}%
  \BibitemOpen
  \bibfield  {author} {\bibinfo {author} {\bibfnamefont {P.}~\bibnamefont
  {Toloueinia}}, \bibinfo {author} {\bibfnamefont {H.}~\bibnamefont {Khassaf}},
  \bibinfo {author} {\bibfnamefont {A.}~\bibnamefont {Shirazi~Amin}}, \bibinfo
  {author} {\bibfnamefont {Z.~M.}\ \bibnamefont {Tobin}}, \bibinfo {author}
  {\bibfnamefont {S.~P.}\ \bibnamefont {Alpay}}, \ and\ \bibinfo {author}
  {\bibfnamefont {S.~L.}\ \bibnamefont {Suib}},\ }\href {\doibase
  https://doi.org/10.1021/acsaem.0c00638} {\bibfield  {journal} {\bibinfo
  {journal} {ACS Applied Energy Materials}\ }\textbf {\bibinfo {volume} {3}},\
  \bibinfo {pages} {8240} (\bibinfo {year} {2020})}\BibitemShut {NoStop}%
\bibitem [{\citenamefont {Montecucco}\ \emph {et~al.}(2021)\citenamefont
  {Montecucco}, \citenamefont {Quadrivi}, \citenamefont {Po},\ and\
  \citenamefont {Grancini}}]{CsPbI3-lifetime}%
  \BibitemOpen
  \bibfield  {author} {\bibinfo {author} {\bibfnamefont {R.}~\bibnamefont
  {Montecucco}}, \bibinfo {author} {\bibfnamefont {E.}~\bibnamefont
  {Quadrivi}}, \bibinfo {author} {\bibfnamefont {R.}~\bibnamefont {Po}}, \ and\
  \bibinfo {author} {\bibfnamefont {G.}~\bibnamefont {Grancini}},\ }\href@noop
  {} {\bibfield  {journal} {\bibinfo  {journal} {Advanced Energy Materials}\
  }\textbf {\bibinfo {volume} {11}},\ \bibinfo {pages} {2100672} (\bibinfo
  {year} {2021})}\BibitemShut {NoStop}%
\bibitem [{\citenamefont {Lewin}\ and\ \citenamefont
  {Smith}(1996)}]{LEWIN199687}%
  \BibitemOpen
  \bibfield  {author} {\bibinfo {author} {\bibfnamefont {J.}~\bibnamefont
  {Lewin}}\ and\ \bibinfo {author} {\bibfnamefont {P.}~\bibnamefont {Smith}},\
  }\href {\doibase https://doi.org/10.1016/S0927-6505(96)00047-3} {\bibfield
  {journal} {\bibinfo  {journal} {Astroparticle Physics}\ }\textbf {\bibinfo
  {volume} {6}},\ \bibinfo {pages} {87} (\bibinfo {year} {1996})}\BibitemShut
  {NoStop}%
\bibitem [{\citenamefont {Catena}\ and\ \citenamefont
  {Ullio}(2010)}]{Catena:2009mf}%
  \BibitemOpen
  \bibfield  {author} {\bibinfo {author} {\bibfnamefont {R.}~\bibnamefont
  {Catena}}\ and\ \bibinfo {author} {\bibfnamefont {P.}~\bibnamefont {Ullio}},\
  }\href {\doibase 10.1088/1475-7516/2010/08/004} {\bibfield  {journal}
  {\bibinfo  {journal} {JCAP}\ }\textbf {\bibinfo {volume} {08}},\ \bibinfo
  {pages} {004} (\bibinfo {year} {2010})},\ \Eprint
  {http://arxiv.org/abs/0907.0018} {arXiv:0907.0018 [astro-ph.CO]} \BibitemShut
  {NoStop}%
\bibitem [{\citenamefont {Salucci}\ \emph {et~al.}(2010)\citenamefont
  {Salucci}, \citenamefont {Nesti}, \citenamefont {Gentile},\ and\
  \citenamefont {Martins}}]{Salucci:2010qr}%
  \BibitemOpen
  \bibfield  {author} {\bibinfo {author} {\bibfnamefont {P.}~\bibnamefont
  {Salucci}}, \bibinfo {author} {\bibfnamefont {F.}~\bibnamefont {Nesti}},
  \bibinfo {author} {\bibfnamefont {G.}~\bibnamefont {Gentile}}, \ and\
  \bibinfo {author} {\bibfnamefont {C.~F.}\ \bibnamefont {Martins}},\ }\href
  {\doibase 10.1051/0004-6361/201014385} {\bibfield  {journal} {\bibinfo
  {journal} {Astron. Astrophys.}\ }\textbf {\bibinfo {volume} {523}},\ \bibinfo
  {pages} {A83} (\bibinfo {year} {2010})},\ \Eprint
  {http://arxiv.org/abs/1003.3101} {arXiv:1003.3101 [astro-ph.GA]} \BibitemShut
  {NoStop}%
\bibitem [{\citenamefont {Knapen}\ \emph
  {et~al.}(2017{\natexlab{b}})\citenamefont {Knapen}, \citenamefont {Lin},\
  and\ \citenamefont {Zurek}}]{Knapen_2017}%
  \BibitemOpen
  \bibfield  {author} {\bibinfo {author} {\bibfnamefont {S.}~\bibnamefont
  {Knapen}}, \bibinfo {author} {\bibfnamefont {T.}~\bibnamefont {Lin}}, \ and\
  \bibinfo {author} {\bibfnamefont {K.~M.}\ \bibnamefont {Zurek}},\ }\href
  {\doibase 10.1103/physrevd.96.115021} {\bibfield  {journal} {\bibinfo
  {journal} {Physical Review D}\ }\textbf {\bibinfo {volume} {96}} (\bibinfo
  {year} {2017}{\natexlab{b}}),\ 10.1103/physrevd.96.115021}\BibitemShut
  {NoStop}%
\bibitem [{\citenamefont {Trickle}\ \emph {et~al.}(2022)\citenamefont
  {Trickle}, \citenamefont {Zhang},\ and\ \citenamefont {Zurek}}]{PhonoDark}%
  \BibitemOpen
  \bibfield  {author} {\bibinfo {author} {\bibfnamefont {T.}~\bibnamefont
  {Trickle}}, \bibinfo {author} {\bibfnamefont {Z.}~\bibnamefont {Zhang}}, \
  and\ \bibinfo {author} {\bibfnamefont {K.~M.}\ \bibnamefont {Zurek}},\ }\href
  {\doibase 10.1103/PhysRevD.105.015001} {\bibfield  {journal} {\bibinfo
  {journal} {Phys. Rev. D}\ }\textbf {\bibinfo {volume} {105}},\ \bibinfo
  {pages} {015001} (\bibinfo {year} {2022})},\ \Eprint
  {http://arxiv.org/abs/2009.13534} {arXiv:2009.13534 [hep-ph]} \BibitemShut
  {NoStop}%
\bibitem [{\citenamefont {Knapen}\ \emph {et~al.}(2022)\citenamefont {Knapen},
  \citenamefont {Kozaczuk},\ and\ \citenamefont {Lin}}]{DarkELF}%
  \BibitemOpen
  \bibfield  {author} {\bibinfo {author} {\bibfnamefont {S.}~\bibnamefont
  {Knapen}}, \bibinfo {author} {\bibfnamefont {J.}~\bibnamefont {Kozaczuk}}, \
  and\ \bibinfo {author} {\bibfnamefont {T.}~\bibnamefont {Lin}},\ }\href
  {\doibase https://doi.org/10.1103/PhysRevD.105.015014} {\bibfield  {journal}
  {\bibinfo  {journal} {Phys. Rev. D}\ }\textbf {\bibinfo {volume} {105}},\
  \bibinfo {pages} {015014} (\bibinfo {year} {2022})}\BibitemShut {NoStop}%
\bibitem [{\citenamefont {Gervais}\ and\ \citenamefont
  {Piriou}(1974)}]{Gervais1974}%
  \BibitemOpen
  \bibfield  {author} {\bibinfo {author} {\bibfnamefont {F.}~\bibnamefont
  {Gervais}}\ and\ \bibinfo {author} {\bibfnamefont {B.}~\bibnamefont
  {Piriou}},\ }\href {\doibase 10.1088/0022-3719/7/13/017} {\bibfield
  {journal} {\bibinfo  {journal} {Journal of Physics C: Solid State Physics}\
  }\textbf {\bibinfo {volume} {7}},\ \bibinfo {pages} {2374} (\bibinfo {year}
  {1974})}\BibitemShut {NoStop}%
\bibitem [{\citenamefont {Zollner}\ \emph {et~al.}(2019)\citenamefont
  {Zollner}, \citenamefont {Paradis}, \citenamefont {Abadizaman},\ and\
  \citenamefont {Samarasingha}}]{ZollerJVST2019}%
  \BibitemOpen
  \bibfield  {author} {\bibinfo {author} {\bibfnamefont {S.}~\bibnamefont
  {Zollner}}, \bibinfo {author} {\bibfnamefont {P.~P.}\ \bibnamefont
  {Paradis}}, \bibinfo {author} {\bibfnamefont {F.}~\bibnamefont {Abadizaman}},
  \ and\ \bibinfo {author} {\bibfnamefont {N.~S.}\ \bibnamefont
  {Samarasingha}},\ }\href {\doibase 10.1116/1.5081055} {\bibfield  {journal}
  {\bibinfo  {journal} {Journal of Vacuum Science \& Technology B}\ }\textbf
  {\bibinfo {volume} {37}},\ \bibinfo {pages} {012904} (\bibinfo {year}
  {2019})},\ \Eprint
  {http://arxiv.org/abs/https://pubs.aip.org/avs/jvb/article-pdf/doi/10.1116/1.5081055/14735213/012904\_1\_online.pdf}
  {https://pubs.aip.org/avs/jvb/article-pdf/doi/10.1116/1.5081055/14735213/012904\_1\_online.pdf}
  \BibitemShut {NoStop}%
\bibitem [{\citenamefont {Fr{\"o}hlich}(1954)}]{Frohlich}%
  \BibitemOpen
  \bibfield  {author} {\bibinfo {author} {\bibfnamefont {H.}~\bibnamefont
  {Fr{\"o}hlich}},\ }\href {\doibase 10.1080/00018735400101213} {\bibfield
  {journal} {\bibinfo  {journal} {Advances in Physics}\ }\textbf {\bibinfo
  {volume} {3}},\ \bibinfo {pages} {325} (\bibinfo {year} {1954})},\ \Eprint
  {http://arxiv.org/abs/https://doi.org/10.1080/00018735400101213}
  {https://doi.org/10.1080/00018735400101213} \BibitemShut {NoStop}%
\bibitem [{\citenamefont {M\"uller}\ and\ \citenamefont
  {Burkard}(1979)}]{PhysRevB.19.3593}%
  \BibitemOpen
  \bibfield  {author} {\bibinfo {author} {\bibfnamefont {K.~A.}\ \bibnamefont
  {M\"uller}}\ and\ \bibinfo {author} {\bibfnamefont {H.}~\bibnamefont
  {Burkard}},\ }\href {\doibase https://doi.org/10.1103/PhysRevB.19.3593}
  {\bibfield  {journal} {\bibinfo  {journal} {Phys. Rev. B}\ }\textbf {\bibinfo
  {volume} {19}},\ \bibinfo {pages} {3593} (\bibinfo {year}
  {1979})}\BibitemShut {NoStop}%
\bibitem [{\citenamefont {Sendner}\ \emph {et~al.}(2016)\citenamefont
  {Sendner}, \citenamefont {Nayak}, \citenamefont {Egger}, \citenamefont
  {Beck}, \citenamefont {Müller}, \citenamefont {Epding}, \citenamefont
  {Kowalsky}, \citenamefont {Kronik}, \citenamefont {Snaith}, \citenamefont
  {Pucci},\ and\ \citenamefont {Lovrinčić}}]{C6MH00275G}%
  \BibitemOpen
  \bibfield  {author} {\bibinfo {author} {\bibfnamefont {M.}~\bibnamefont
  {Sendner}}, \bibinfo {author} {\bibfnamefont {P.~K.}\ \bibnamefont {Nayak}},
  \bibinfo {author} {\bibfnamefont {D.~A.}\ \bibnamefont {Egger}}, \bibinfo
  {author} {\bibfnamefont {S.}~\bibnamefont {Beck}}, \bibinfo {author}
  {\bibfnamefont {C.}~\bibnamefont {Müller}}, \bibinfo {author} {\bibfnamefont
  {B.}~\bibnamefont {Epding}}, \bibinfo {author} {\bibfnamefont
  {W.}~\bibnamefont {Kowalsky}}, \bibinfo {author} {\bibfnamefont
  {L.}~\bibnamefont {Kronik}}, \bibinfo {author} {\bibfnamefont {H.~J.}\
  \bibnamefont {Snaith}}, \bibinfo {author} {\bibfnamefont {A.}~\bibnamefont
  {Pucci}}, \ and\ \bibinfo {author} {\bibfnamefont {R.}~\bibnamefont
  {Lovrinčić}},\ }\href {\doibase 10.1039/C6MH00275G} {\bibfield  {journal}
  {\bibinfo  {journal} {Mater. Horiz.}\ }\textbf {\bibinfo {volume} {3}},\
  \bibinfo {pages} {613} (\bibinfo {year} {2016})}\BibitemShut {NoStop}%
\bibitem [{\citenamefont {Maeng}\ \emph {et~al.}(2023)\citenamefont {Maeng},
  \citenamefont {Chen}, \citenamefont {Lee}, \citenamefont {Wang},
  \citenamefont {Kwon},\ and\ \citenamefont
  {Jung}}]{GammaCsPbI3_Lattice2_MatPhyToday_2023}%
  \BibitemOpen
  \bibfield  {author} {\bibinfo {author} {\bibfnamefont {I.}~\bibnamefont
  {Maeng}}, \bibinfo {author} {\bibfnamefont {S.}~\bibnamefont {Chen}},
  \bibinfo {author} {\bibfnamefont {S.}~\bibnamefont {Lee}}, \bibinfo {author}
  {\bibfnamefont {S.}~\bibnamefont {Wang}}, \bibinfo {author} {\bibfnamefont
  {Y.-K.}\ \bibnamefont {Kwon}}, \ and\ \bibinfo {author} {\bibfnamefont
  {M.-C.}\ \bibnamefont {Jung}},\ }\href {\doibase
  https://doi.org/10.1016/j.mtphys.2022.100960} {\bibfield  {journal} {\bibinfo
   {journal} {Materials Today Physics}\ }\textbf {\bibinfo {volume} {30}},\
  \bibinfo {pages} {100960} (\bibinfo {year} {2023})}\BibitemShut {NoStop}%
\bibitem [{\citenamefont {Fung}\ \emph {et~al.}(2024)\citenamefont {Fung},
  \citenamefont {Heeba}, \citenamefont {Liu}, \citenamefont {Muralidharan},
  \citenamefont {Schutz},\ and\ \citenamefont {Vincent}}]{Stellar1}%
  \BibitemOpen
  \bibfield  {author} {\bibinfo {author} {\bibfnamefont {A.}~\bibnamefont
  {Fung}}, \bibinfo {author} {\bibfnamefont {S.}~\bibnamefont {Heeba}},
  \bibinfo {author} {\bibfnamefont {Q.}~\bibnamefont {Liu}}, \bibinfo {author}
  {\bibfnamefont {V.}~\bibnamefont {Muralidharan}}, \bibinfo {author}
  {\bibfnamefont {K.}~\bibnamefont {Schutz}}, \ and\ \bibinfo {author}
  {\bibfnamefont {A.~C.}\ \bibnamefont {Vincent}},\ }\href {\doibase
  https://doi.org/10.1103/PhysRevD.109.083011} {\bibfield  {journal} {\bibinfo
  {journal} {Phys. Rev. D}\ }\textbf {\bibinfo {volume} {109}},\ \bibinfo
  {pages} {083011} (\bibinfo {year} {2024})},\ \Eprint
  {http://arxiv.org/abs/2309.06465} {arXiv:2309.06465 [hep-ph]} \BibitemShut
  {NoStop}%
\bibitem [{\citenamefont {Knab}\ \emph {et~al.}(2013)\citenamefont {Knab},
  \citenamefont {Adam}, \citenamefont {Shaner}, \citenamefont {Starmans},\ and\
  \citenamefont {Planken}}]{CsI_Frequencies}%
  \BibitemOpen
  \bibfield  {author} {\bibinfo {author} {\bibfnamefont {J.~R.}\ \bibnamefont
  {Knab}}, \bibinfo {author} {\bibfnamefont {A.~J.~L.}\ \bibnamefont {Adam}},
  \bibinfo {author} {\bibfnamefont {E.}~\bibnamefont {Shaner}}, \bibinfo
  {author} {\bibfnamefont {H.~J. A.~J.}\ \bibnamefont {Starmans}}, \ and\
  \bibinfo {author} {\bibfnamefont {P.~C.~M.}\ \bibnamefont {Planken}},\ }\href
  {\doibase 10.1364/OE.21.001101} {\bibfield  {journal} {\bibinfo  {journal}
  {Opt. Express}\ }\textbf {\bibinfo {volume} {21}},\ \bibinfo {pages} {1101}
  (\bibinfo {year} {2013})}\BibitemShut {NoStop}%
\bibitem [{\citenamefont {Kaplunov}\ \emph {et~al.}(2021)\citenamefont
  {Kaplunov}, \citenamefont {Kropotov}, \citenamefont {Rogalin},\ and\
  \citenamefont {Shakhmin}}]{kaplunov2021transmittance}%
  \BibitemOpen
  \bibfield  {author} {\bibinfo {author} {\bibfnamefont {I.~A.}\ \bibnamefont
  {Kaplunov}}, \bibinfo {author} {\bibfnamefont {G.~I.}\ \bibnamefont
  {Kropotov}}, \bibinfo {author} {\bibfnamefont {V.~E.}\ \bibnamefont
  {Rogalin}}, \ and\ \bibinfo {author} {\bibfnamefont {A.~A.}\ \bibnamefont
  {Shakhmin}},\ }\href {\doibase
  https://doi.org/10.21883/OS.2020.10.50017.128-20} {\bibfield  {journal}
  {\bibinfo  {journal} {Optics and Spectroscopy}\ }\textbf {\bibinfo {volume}
  {129}},\ \bibinfo {pages} {775} (\bibinfo {year} {2021})}\BibitemShut
  {NoStop}%
\bibitem [{\citenamefont {DeRocco}\ \emph {et~al.}(2020)\citenamefont
  {DeRocco}, \citenamefont {Graham},\ and\ \citenamefont
  {Rajendran}}]{Stellar2}%
  \BibitemOpen
  \bibfield  {author} {\bibinfo {author} {\bibfnamefont {W.}~\bibnamefont
  {DeRocco}}, \bibinfo {author} {\bibfnamefont {P.~W.}\ \bibnamefont {Graham}},
  \ and\ \bibinfo {author} {\bibfnamefont {S.}~\bibnamefont {Rajendran}},\
  }\href {\doibase https://doi.org/10.1103/PhysRevD.102.075015} {\bibfield
  {journal} {\bibinfo  {journal} {Phys. Rev. D}\ }\textbf {\bibinfo {volume}
  {102}},\ \bibinfo {pages} {075015} (\bibinfo {year} {2020})}\BibitemShut
  {NoStop}%
\bibitem [{\citenamefont {Davidson}\ \emph {et~al.}(2000)\citenamefont
  {Davidson}, \citenamefont {Hannestad},\ and\ \citenamefont
  {Raffelt}}]{Stellar3}%
  \BibitemOpen
  \bibfield  {author} {\bibinfo {author} {\bibfnamefont {S.}~\bibnamefont
  {Davidson}}, \bibinfo {author} {\bibfnamefont {S.}~\bibnamefont {Hannestad}},
  \ and\ \bibinfo {author} {\bibfnamefont {G.}~\bibnamefont {Raffelt}},\ }\href
  {\doibase https://doi.org/10.1088/1126-6708/2000/05/003} {\bibfield
  {journal} {\bibinfo  {journal} {JHEP}\ }\textbf {\bibinfo {volume} {05}},\
  \bibinfo {pages} {003} (\bibinfo {year} {2000})},\ \Eprint
  {http://arxiv.org/abs/hep-ph/0001179} {arXiv:hep-ph/0001179} \BibitemShut
  {NoStop}%
\bibitem [{\citenamefont {Giannozzi}\ \emph {et~al.}(2009)\citenamefont
  {Giannozzi}, \citenamefont {Baroni}, \citenamefont {Bonini}, \citenamefont
  {Calandra}, \citenamefont {Car}, \citenamefont {Cavazzoni}, \citenamefont
  {Ceresoli}, \citenamefont {Chiarotti}, \citenamefont {Cococcioni},
  \citenamefont {Dabo}, \citenamefont {{Dal Corso}}, \citenamefont
  {de~Gironcoli}, \citenamefont {Fabris}, \citenamefont {Fratesi},
  \citenamefont {Gebauer}, \citenamefont {Gerstmann}, \citenamefont
  {Gougoussis}, \citenamefont {Kokalj}, \citenamefont {Lazzeri}, \citenamefont
  {Martin-Samos}, \citenamefont {Marzari}, \citenamefont {Mauri}, \citenamefont
  {Mazzarello}, \citenamefont {Paolini}, \citenamefont {Pasquarello},
  \citenamefont {Paulatto}, \citenamefont {Sbraccia}, \citenamefont {Scandolo},
  \citenamefont {Sclauzero}, \citenamefont {Seitsonen}, \citenamefont
  {Smogunov}, \citenamefont {Umari},\ and\ \citenamefont
  {Wentzcovitch}}]{QE-2009}%
  \BibitemOpen
  \bibfield  {author} {\bibinfo {author} {\bibfnamefont {P.}~\bibnamefont
  {Giannozzi}}, \bibinfo {author} {\bibfnamefont {S.}~\bibnamefont {Baroni}},
  \bibinfo {author} {\bibfnamefont {N.}~\bibnamefont {Bonini}}, \bibinfo
  {author} {\bibfnamefont {M.}~\bibnamefont {Calandra}}, \bibinfo {author}
  {\bibfnamefont {R.}~\bibnamefont {Car}}, \bibinfo {author} {\bibfnamefont
  {C.}~\bibnamefont {Cavazzoni}}, \bibinfo {author} {\bibfnamefont
  {D.}~\bibnamefont {Ceresoli}}, \bibinfo {author} {\bibfnamefont {G.~L.}\
  \bibnamefont {Chiarotti}}, \bibinfo {author} {\bibfnamefont {M.}~\bibnamefont
  {Cococcioni}}, \bibinfo {author} {\bibfnamefont {I.}~\bibnamefont {Dabo}},
  \bibinfo {author} {\bibfnamefont {A.}~\bibnamefont {{Dal Corso}}}, \bibinfo
  {author} {\bibfnamefont {S.}~\bibnamefont {de~Gironcoli}}, \bibinfo {author}
  {\bibfnamefont {S.}~\bibnamefont {Fabris}}, \bibinfo {author} {\bibfnamefont
  {G.}~\bibnamefont {Fratesi}}, \bibinfo {author} {\bibfnamefont
  {R.}~\bibnamefont {Gebauer}}, \bibinfo {author} {\bibfnamefont
  {U.}~\bibnamefont {Gerstmann}}, \bibinfo {author} {\bibfnamefont
  {C.}~\bibnamefont {Gougoussis}}, \bibinfo {author} {\bibfnamefont
  {A.}~\bibnamefont {Kokalj}}, \bibinfo {author} {\bibfnamefont
  {M.}~\bibnamefont {Lazzeri}}, \bibinfo {author} {\bibfnamefont
  {L.}~\bibnamefont {Martin-Samos}}, \bibinfo {author} {\bibfnamefont
  {N.}~\bibnamefont {Marzari}}, \bibinfo {author} {\bibfnamefont
  {F.}~\bibnamefont {Mauri}}, \bibinfo {author} {\bibfnamefont
  {R.}~\bibnamefont {Mazzarello}}, \bibinfo {author} {\bibfnamefont
  {S.}~\bibnamefont {Paolini}}, \bibinfo {author} {\bibfnamefont
  {A.}~\bibnamefont {Pasquarello}}, \bibinfo {author} {\bibfnamefont
  {L.}~\bibnamefont {Paulatto}}, \bibinfo {author} {\bibfnamefont
  {C.}~\bibnamefont {Sbraccia}}, \bibinfo {author} {\bibfnamefont
  {S.}~\bibnamefont {Scandolo}}, \bibinfo {author} {\bibfnamefont
  {G.}~\bibnamefont {Sclauzero}}, \bibinfo {author} {\bibfnamefont {A.~P.}\
  \bibnamefont {Seitsonen}}, \bibinfo {author} {\bibfnamefont {A.}~\bibnamefont
  {Smogunov}}, \bibinfo {author} {\bibfnamefont {P.}~\bibnamefont {Umari}}, \
  and\ \bibinfo {author} {\bibfnamefont {R.~M.}\ \bibnamefont {Wentzcovitch}},\
  }\href {http://www.quantum-espresso.org} {\bibfield  {journal} {\bibinfo
  {journal} {Journal of Physics: Condensed Matter}\ }\textbf {\bibinfo {volume}
  {21}},\ \bibinfo {pages} {395502 (19pp)} (\bibinfo {year}
  {2009})}\BibitemShut {NoStop}%
\bibitem [{\citenamefont {Giannozzi}\ \emph {et~al.}(2017)\citenamefont
  {Giannozzi}, \citenamefont {Andreussi}, \citenamefont {Brumme}, \citenamefont
  {Bunau}, \citenamefont {Nardelli}, \citenamefont {Calandra}, \citenamefont
  {Car}, \citenamefont {Cavazzoni}, \citenamefont {Ceresoli}, \citenamefont
  {Cococcioni}, \citenamefont {Colonna}, \citenamefont {Carnimeo},
  \citenamefont {Corso}, \citenamefont {de~Gironcoli}, \citenamefont {Delugas},
  \citenamefont {Jr}, \citenamefont {Ferretti}, \citenamefont {Floris},
  \citenamefont {Fratesi}, \citenamefont {Fugallo}, \citenamefont {Gebauer},
  \citenamefont {Gerstmann}, \citenamefont {Giustino}, \citenamefont {Gorni},
  \citenamefont {Jia}, \citenamefont {Kawamura}, \citenamefont {Ko},
  \citenamefont {Kokalj}, \citenamefont {Küçükbenli}, \citenamefont
  {Lazzeri}, \citenamefont {Marsili}, \citenamefont {Marzari}, \citenamefont
  {Mauri}, \citenamefont {Nguyen}, \citenamefont {Nguyen}, \citenamefont {de-la
  Roza}, \citenamefont {Paulatto}, \citenamefont {Poncé}, \citenamefont
  {Rocca}, \citenamefont {Sabatini}, \citenamefont {Santra}, \citenamefont
  {Schlipf}, \citenamefont {Seitsonen}, \citenamefont {Smogunov}, \citenamefont
  {Timrov}, \citenamefont {Thonhauser}, \citenamefont {Umari}, \citenamefont
  {Vast}, \citenamefont {Wu},\ and\ \citenamefont {Baroni}}]{QE-2017}%
  \BibitemOpen
  \bibfield  {author} {\bibinfo {author} {\bibfnamefont {P.}~\bibnamefont
  {Giannozzi}}, \bibinfo {author} {\bibfnamefont {O.}~\bibnamefont
  {Andreussi}}, \bibinfo {author} {\bibfnamefont {T.}~\bibnamefont {Brumme}},
  \bibinfo {author} {\bibfnamefont {O.}~\bibnamefont {Bunau}}, \bibinfo
  {author} {\bibfnamefont {M.~B.}\ \bibnamefont {Nardelli}}, \bibinfo {author}
  {\bibfnamefont {M.}~\bibnamefont {Calandra}}, \bibinfo {author}
  {\bibfnamefont {R.}~\bibnamefont {Car}}, \bibinfo {author} {\bibfnamefont
  {C.}~\bibnamefont {Cavazzoni}}, \bibinfo {author} {\bibfnamefont
  {D.}~\bibnamefont {Ceresoli}}, \bibinfo {author} {\bibfnamefont
  {M.}~\bibnamefont {Cococcioni}}, \bibinfo {author} {\bibfnamefont
  {N.}~\bibnamefont {Colonna}}, \bibinfo {author} {\bibfnamefont
  {I.}~\bibnamefont {Carnimeo}}, \bibinfo {author} {\bibfnamefont {A.~D.}\
  \bibnamefont {Corso}}, \bibinfo {author} {\bibfnamefont {S.}~\bibnamefont
  {de~Gironcoli}}, \bibinfo {author} {\bibfnamefont {P.}~\bibnamefont
  {Delugas}}, \bibinfo {author} {\bibfnamefont {R.~A.~D.}\ \bibnamefont {Jr}},
  \bibinfo {author} {\bibfnamefont {A.}~\bibnamefont {Ferretti}}, \bibinfo
  {author} {\bibfnamefont {A.}~\bibnamefont {Floris}}, \bibinfo {author}
  {\bibfnamefont {G.}~\bibnamefont {Fratesi}}, \bibinfo {author} {\bibfnamefont
  {G.}~\bibnamefont {Fugallo}}, \bibinfo {author} {\bibfnamefont
  {R.}~\bibnamefont {Gebauer}}, \bibinfo {author} {\bibfnamefont
  {U.}~\bibnamefont {Gerstmann}}, \bibinfo {author} {\bibfnamefont
  {F.}~\bibnamefont {Giustino}}, \bibinfo {author} {\bibfnamefont
  {T.}~\bibnamefont {Gorni}}, \bibinfo {author} {\bibfnamefont
  {J.}~\bibnamefont {Jia}}, \bibinfo {author} {\bibfnamefont {M.}~\bibnamefont
  {Kawamura}}, \bibinfo {author} {\bibfnamefont {H.-Y.}\ \bibnamefont {Ko}},
  \bibinfo {author} {\bibfnamefont {A.}~\bibnamefont {Kokalj}}, \bibinfo
  {author} {\bibfnamefont {E.}~\bibnamefont {Küçükbenli}}, \bibinfo {author}
  {\bibfnamefont {M.}~\bibnamefont {Lazzeri}}, \bibinfo {author} {\bibfnamefont
  {M.}~\bibnamefont {Marsili}}, \bibinfo {author} {\bibfnamefont
  {N.}~\bibnamefont {Marzari}}, \bibinfo {author} {\bibfnamefont
  {F.}~\bibnamefont {Mauri}}, \bibinfo {author} {\bibfnamefont {N.~L.}\
  \bibnamefont {Nguyen}}, \bibinfo {author} {\bibfnamefont {H.-V.}\
  \bibnamefont {Nguyen}}, \bibinfo {author} {\bibfnamefont {A.~O.}\
  \bibnamefont {de-la Roza}}, \bibinfo {author} {\bibfnamefont
  {L.}~\bibnamefont {Paulatto}}, \bibinfo {author} {\bibfnamefont
  {S.}~\bibnamefont {Poncé}}, \bibinfo {author} {\bibfnamefont
  {D.}~\bibnamefont {Rocca}}, \bibinfo {author} {\bibfnamefont
  {R.}~\bibnamefont {Sabatini}}, \bibinfo {author} {\bibfnamefont
  {B.}~\bibnamefont {Santra}}, \bibinfo {author} {\bibfnamefont
  {M.}~\bibnamefont {Schlipf}}, \bibinfo {author} {\bibfnamefont {A.~P.}\
  \bibnamefont {Seitsonen}}, \bibinfo {author} {\bibfnamefont {A.}~\bibnamefont
  {Smogunov}}, \bibinfo {author} {\bibfnamefont {I.}~\bibnamefont {Timrov}},
  \bibinfo {author} {\bibfnamefont {T.}~\bibnamefont {Thonhauser}}, \bibinfo
  {author} {\bibfnamefont {P.}~\bibnamefont {Umari}}, \bibinfo {author}
  {\bibfnamefont {N.}~\bibnamefont {Vast}}, \bibinfo {author} {\bibfnamefont
  {X.}~\bibnamefont {Wu}}, \ and\ \bibinfo {author} {\bibfnamefont
  {S.}~\bibnamefont {Baroni}},\ }\href
  {http://stacks.iop.org/0953-8984/29/i=46/a=465901} {\bibfield  {journal}
  {\bibinfo  {journal} {Journal of Physics: Condensed Matter}\ }\textbf
  {\bibinfo {volume} {29}},\ \bibinfo {pages} {465901} (\bibinfo {year}
  {2017})}\BibitemShut {NoStop}%
\bibitem [{\citenamefont {Marronnier}\ \emph {et~al.}(2018)\citenamefont
  {Marronnier}, \citenamefont {Roma}, \citenamefont {Boyer-Richard},
  \citenamefont {Pedesseau}, \citenamefont {Jancu}, \citenamefont
  {Bonnassieux}, \citenamefont {Katan}, \citenamefont {Stoumpos}, \citenamefont
  {Kanatzidis},\ and\ \citenamefont
  {Even}}]{GammaCsPbI3_Marronier_ACSNano_2018}%
  \BibitemOpen
  \bibfield  {author} {\bibinfo {author} {\bibfnamefont {A.}~\bibnamefont
  {Marronnier}}, \bibinfo {author} {\bibfnamefont {G.}~\bibnamefont {Roma}},
  \bibinfo {author} {\bibfnamefont {S.}~\bibnamefont {Boyer-Richard}}, \bibinfo
  {author} {\bibfnamefont {L.}~\bibnamefont {Pedesseau}}, \bibinfo {author}
  {\bibfnamefont {J.-M.}\ \bibnamefont {Jancu}}, \bibinfo {author}
  {\bibfnamefont {Y.}~\bibnamefont {Bonnassieux}}, \bibinfo {author}
  {\bibfnamefont {C.}~\bibnamefont {Katan}}, \bibinfo {author} {\bibfnamefont
  {C.~C.}\ \bibnamefont {Stoumpos}}, \bibinfo {author} {\bibfnamefont {M.~G.}\
  \bibnamefont {Kanatzidis}}, \ and\ \bibinfo {author} {\bibfnamefont
  {J.}~\bibnamefont {Even}},\ }\href {\doibase 10.1021/acsnano.8b00267}
  {\bibfield  {journal} {\bibinfo  {journal} {ACS Nano}\ }\textbf {\bibinfo
  {volume} {12}},\ \bibinfo {pages} {3477} (\bibinfo {year} {2018})},\ \bibinfo
  {note} {pMID: 29565559},\ \Eprint
  {http://arxiv.org/abs/https://doi.org/10.1021/acsnano.8b00267}
  {https://doi.org/10.1021/acsnano.8b00267} \BibitemShut {NoStop}%
\bibitem [{\citenamefont {Straus}\ \emph {et~al.}(2019)\citenamefont {Straus},
  \citenamefont {Guo},\ and\ \citenamefont {Cava}}]{GAMMACSPBI3}%
  \BibitemOpen
  \bibfield  {author} {\bibinfo {author} {\bibfnamefont {D.~B.}\ \bibnamefont
  {Straus}}, \bibinfo {author} {\bibfnamefont {S.}~\bibnamefont {Guo}}, \ and\
  \bibinfo {author} {\bibfnamefont {R.~J.}\ \bibnamefont {Cava}},\ }\href
  {\doibase https://doi.org/10.1021/jacs.9b06055} {\bibfield  {journal}
  {\bibinfo  {journal} {Journal of the American Chemical Society}\ }\textbf
  {\bibinfo {volume} {141}},\ \bibinfo {pages} {11435} (\bibinfo {year}
  {2019})}\BibitemShut {NoStop}%
\bibitem [{\citenamefont {Glazer}(1975)}]{GLAZERPEROVSKITES}%
  \BibitemOpen
  \bibfield  {author} {\bibinfo {author} {\bibfnamefont {A.}~\bibnamefont
  {Glazer}},\ }\href {\doibase http://dx.doi.org/10.1107/S0567739475001635}
  {\bibfield  {journal} {\bibinfo  {journal} {Acta Crystallographica Section A:
  Crystal Physics, Diffraction, Theoretical and General Crystallography}\
  }\textbf {\bibinfo {volume} {31}},\ \bibinfo {pages} {756} (\bibinfo {year}
  {1975})}\BibitemShut {NoStop}%
\bibitem [{\citenamefont {Monkhorst}\ and\ \citenamefont
  {Pack}(1976)}]{MPgrid}%
  \BibitemOpen
  \bibfield  {author} {\bibinfo {author} {\bibfnamefont {H.~J.}\ \bibnamefont
  {Monkhorst}}\ and\ \bibinfo {author} {\bibfnamefont {J.~D.}\ \bibnamefont
  {Pack}},\ }\href {\doibase https://doi.org/10.1103/PhysRevB.13.5188}
  {\bibfield  {journal} {\bibinfo  {journal} {Phys. Rev. B}\ }\textbf {\bibinfo
  {volume} {13}},\ \bibinfo {pages} {5188} (\bibinfo {year}
  {1976})}\BibitemShut {NoStop}%
\bibitem [{\citenamefont {Togo}\ \emph {et~al.}(2023)\citenamefont {Togo},
  \citenamefont {Chaput}, \citenamefont {Tadano},\ and\ \citenamefont
  {Tanaka}}]{phonopy-phono3py-JPCM}%
  \BibitemOpen
  \bibfield  {author} {\bibinfo {author} {\bibfnamefont {A.}~\bibnamefont
  {Togo}}, \bibinfo {author} {\bibfnamefont {L.}~\bibnamefont {Chaput}},
  \bibinfo {author} {\bibfnamefont {T.}~\bibnamefont {Tadano}}, \ and\ \bibinfo
  {author} {\bibfnamefont {I.}~\bibnamefont {Tanaka}},\ }\href {\doibase
  https://doi.org/10.1088/1361-648X/acd831} {\bibfield  {journal} {\bibinfo
  {journal} {J. Phys. Condens. Matter}\ }\textbf {\bibinfo {volume} {35}},\
  \bibinfo {pages} {353001} (\bibinfo {year} {2023})}\BibitemShut {NoStop}%
\bibitem [{\citenamefont {Togo}(2023)}]{phonopy-phono3py-JPSJ}%
  \BibitemOpen
  \bibfield  {author} {\bibinfo {author} {\bibfnamefont {A.}~\bibnamefont
  {Togo}},\ }\href {\doibase https://doi.org/10.7566/JPSJ.92.012001} {\bibfield
   {journal} {\bibinfo  {journal} {J. Phys. Soc. Jpn.}\ }\textbf {\bibinfo
  {volume} {92}},\ \bibinfo {pages} {012001} (\bibinfo {year}
  {2023})}\BibitemShut {NoStop}%
\bibitem [{\citenamefont {Pick}\ \emph {et~al.}(1970)\citenamefont {Pick},
  \citenamefont {Cohen},\ and\ \citenamefont {Martin}}]{nac1}%
  \BibitemOpen
  \bibfield  {author} {\bibinfo {author} {\bibfnamefont {R.~M.}\ \bibnamefont
  {Pick}}, \bibinfo {author} {\bibfnamefont {M.~H.}\ \bibnamefont {Cohen}}, \
  and\ \bibinfo {author} {\bibfnamefont {R.~M.}\ \bibnamefont {Martin}},\
  }\href {\doibase https://doi.org/10.1103/PhysRevB.1.910} {\bibfield
  {journal} {\bibinfo  {journal} {Phys. Rev. B}\ }\textbf {\bibinfo {volume}
  {1}},\ \bibinfo {pages} {910} (\bibinfo {year} {1970})},\ \bibinfo {note}
  {\url{https://doi.org/10.1103/PhysRevB.1.910}}\BibitemShut {NoStop}%
\bibitem [{\citenamefont {Giannozzi}\ \emph {et~al.}(1991)\citenamefont
  {Giannozzi}, \citenamefont {de~Gironcoli}, \citenamefont {Pavone},\ and\
  \citenamefont {Baroni}}]{nac2}%
  \BibitemOpen
  \bibfield  {author} {\bibinfo {author} {\bibfnamefont {P.}~\bibnamefont
  {Giannozzi}}, \bibinfo {author} {\bibfnamefont {S.}~\bibnamefont
  {de~Gironcoli}}, \bibinfo {author} {\bibfnamefont {P.}~\bibnamefont
  {Pavone}}, \ and\ \bibinfo {author} {\bibfnamefont {S.}~\bibnamefont
  {Baroni}},\ }\href {\doibase https://doi.org/10.1103/PhysRevB.43.7231}
  {\bibfield  {journal} {\bibinfo  {journal} {Phys. Rev. B}\ }\textbf {\bibinfo
  {volume} {43}},\ \bibinfo {pages} {7231} (\bibinfo {year}
  {1991})}\BibitemShut {NoStop}%
\bibitem [{\citenamefont {Gonze}\ \emph {et~al.}(1994)\citenamefont {Gonze},
  \citenamefont {Charlier}, \citenamefont {Allan},\ and\ \citenamefont
  {Teter}}]{nac3}%
  \BibitemOpen
  \bibfield  {author} {\bibinfo {author} {\bibfnamefont {X.}~\bibnamefont
  {Gonze}}, \bibinfo {author} {\bibfnamefont {J.-C.}\ \bibnamefont {Charlier}},
  \bibinfo {author} {\bibfnamefont {D.}~\bibnamefont {Allan}}, \ and\ \bibinfo
  {author} {\bibfnamefont {M.}~\bibnamefont {Teter}},\ }\href {\doibase
  https://doi.org/10.1103/PhysRevB.50.13035} {\bibfield  {journal} {\bibinfo
  {journal} {Phys. Rev. B}\ }\textbf {\bibinfo {volume} {50}},\ \bibinfo
  {pages} {13035} (\bibinfo {year} {1994})}\BibitemShut {NoStop}%
\bibitem [{\citenamefont {Knapen}\ \emph {et~al.}(2018)\citenamefont {Knapen},
  \citenamefont {Lin}, \citenamefont {Pyle},\ and\ \citenamefont
  {Zurek}}]{Knapen_2018}%
  \BibitemOpen
  \bibfield  {author} {\bibinfo {author} {\bibfnamefont {S.}~\bibnamefont
  {Knapen}}, \bibinfo {author} {\bibfnamefont {T.}~\bibnamefont {Lin}},
  \bibinfo {author} {\bibfnamefont {M.}~\bibnamefont {Pyle}}, \ and\ \bibinfo
  {author} {\bibfnamefont {K.~M.}\ \bibnamefont {Zurek}},\ }\href {\doibase
  https://doi.org/10.1016/j.physletb.2018.08.064} {\bibfield  {journal}
  {\bibinfo  {journal} {Physics Letters B}\ }\textbf {\bibinfo {volume}
  {785}},\ \bibinfo {pages} {386–390} (\bibinfo {year} {2018})}\BibitemShut
  {NoStop}%
\bibitem [{\citenamefont {An}\ \emph {et~al.}(2015)\citenamefont {An},
  \citenamefont {Pospelov}, \citenamefont {Pradler},\ and\ \citenamefont
  {Ritz}}]{An_2015}%
  \BibitemOpen
  \bibfield  {author} {\bibinfo {author} {\bibfnamefont {H.}~\bibnamefont
  {An}}, \bibinfo {author} {\bibfnamefont {M.}~\bibnamefont {Pospelov}},
  \bibinfo {author} {\bibfnamefont {J.}~\bibnamefont {Pradler}}, \ and\
  \bibinfo {author} {\bibfnamefont {A.}~\bibnamefont {Ritz}},\ }\href {\doibase
  https://doi.org/10.1016/j.physletb.2015.06.018} {\bibfield  {journal}
  {\bibinfo  {journal} {Physics Letters B}\ }\textbf {\bibinfo {volume}
  {747}},\ \bibinfo {pages} {331–338} (\bibinfo {year} {2015})}\BibitemShut
  {NoStop}%
\bibitem [{\citenamefont {Hochberg}\ \emph {et~al.}(2016)\citenamefont
  {Hochberg}, \citenamefont {Lin},\ and\ \citenamefont
  {Zurek}}]{Hochberg_2016}%
  \BibitemOpen
  \bibfield  {author} {\bibinfo {author} {\bibfnamefont {Y.}~\bibnamefont
  {Hochberg}}, \bibinfo {author} {\bibfnamefont {T.}~\bibnamefont {Lin}}, \
  and\ \bibinfo {author} {\bibfnamefont {K.~M.}\ \bibnamefont {Zurek}},\ }\href
  {\doibase 10.1103/physrevd.94.015019} {\bibfield  {journal} {\bibinfo
  {journal} {Physical Review D}\ }\textbf {\bibinfo {volume} {94}} (\bibinfo
  {year} {2016}),\ 10.1103/physrevd.94.015019}\BibitemShut {NoStop}%
\bibitem [{\citenamefont {Trots}\ and\ \citenamefont
  {Myagkota}(2008)}]{BRIDGMANGROWTHCSPBI3}%
  \BibitemOpen
  \bibfield  {author} {\bibinfo {author} {\bibfnamefont {D.}~\bibnamefont
  {Trots}}\ and\ \bibinfo {author} {\bibfnamefont {S.}~\bibnamefont
  {Myagkota}},\ }\href {\doibase https://doi.org/10.1016/j.jpcs.2008.05.007}
  {\bibfield  {journal} {\bibinfo  {journal} {Journal of Physics and Chemistry
  of Solids}\ }\textbf {\bibinfo {volume} {69}},\ \bibinfo {pages} {2520}
  (\bibinfo {year} {2008})}\BibitemShut {NoStop}%
\bibitem [{\citenamefont {Zhang}\ \emph {et~al.}(2018)\citenamefont {Zhang},
  \citenamefont {Xiao}, \citenamefont {Dong},\ and\ \citenamefont
  {Xu}}]{CSPBI3AQUEOUS}%
  \BibitemOpen
  \bibfield  {author} {\bibinfo {author} {\bibfnamefont {B.-B.}\ \bibnamefont
  {Zhang}}, \bibinfo {author} {\bibfnamefont {B.}~\bibnamefont {Xiao}},
  \bibinfo {author} {\bibfnamefont {S.}~\bibnamefont {Dong}}, \ and\ \bibinfo
  {author} {\bibfnamefont {Y.}~\bibnamefont {Xu}},\ }\href {\doibase
  https://doi.org/10.1016/j.jcrysgro.2018.05.027} {\bibfield  {journal}
  {\bibinfo  {journal} {Journal of Crystal Growth}\ }\textbf {\bibinfo {volume}
  {498}},\ \bibinfo {pages} {1} (\bibinfo {year} {2018})}\BibitemShut {NoStop}%
\bibitem [{\citenamefont {Yang}\ \emph {et~al.}(2002)\citenamefont {Yang},
  \citenamefont {Liao}, \citenamefont {Shen}, \citenamefont {Shao},
  \citenamefont {Ni},\ and\ \citenamefont {Yin}}]{BRIDGMANGROSSO}%
  \BibitemOpen
  \bibfield  {author} {\bibinfo {author} {\bibfnamefont {P.}~\bibnamefont
  {Yang}}, \bibinfo {author} {\bibfnamefont {J.}~\bibnamefont {Liao}}, \bibinfo
  {author} {\bibfnamefont {B.}~\bibnamefont {Shen}}, \bibinfo {author}
  {\bibfnamefont {P.}~\bibnamefont {Shao}}, \bibinfo {author} {\bibfnamefont
  {H.}~\bibnamefont {Ni}}, \ and\ \bibinfo {author} {\bibfnamefont
  {Z.}~\bibnamefont {Yin}},\ }\href {\doibase
  https://doi.org/10.1016/S0022-0248(01)02385-5} {\bibfield  {journal}
  {\bibinfo  {journal} {Journal of Crystal Growth}\ }\textbf {\bibinfo {volume}
  {236}},\ \bibinfo {pages} {589} (\bibinfo {year} {2002})}\BibitemShut
  {NoStop}%
\bibitem [{\citenamefont {Doyle}\ \emph {et~al.}(2008)\citenamefont {Doyle},
  \citenamefont {Mauskopf}, \citenamefont {Naylon}, \citenamefont {Porch},\
  and\ \citenamefont {Duncombe}}]{DOYLEKID}%
  \BibitemOpen
  \bibfield  {author} {\bibinfo {author} {\bibfnamefont {S.}~\bibnamefont
  {Doyle}}, \bibinfo {author} {\bibfnamefont {P.}~\bibnamefont {Mauskopf}},
  \bibinfo {author} {\bibfnamefont {J.}~\bibnamefont {Naylon}}, \bibinfo
  {author} {\bibfnamefont {A.}~\bibnamefont {Porch}}, \ and\ \bibinfo {author}
  {\bibfnamefont {C.}~\bibnamefont {Duncombe}},\ }\href {\doibase
  https://doi.org/10.1007/s10909-007-9685-2} {\bibfield  {journal} {\bibinfo
  {journal} {Journal of Low Temperature Physics}\ }\textbf {\bibinfo {volume}
  {151}},\ \bibinfo {pages} {530} (\bibinfo {year} {2008})}\BibitemShut
  {NoStop}%
\bibitem [{\citenamefont {Battistelli}\ \emph {et~al.}(2015)\citenamefont
  {Battistelli}, \citenamefont {Bellini}, \citenamefont {Bucci}, \citenamefont
  {Calvo}, \citenamefont {Cardani}, \citenamefont {Casali}, \citenamefont
  {Castellano}, \citenamefont {Colantoni}, \citenamefont {Coppolecchia},
  \citenamefont {Cosmelli} \emph {et~al.}}]{CALDER}%
  \BibitemOpen
  \bibfield  {author} {\bibinfo {author} {\bibfnamefont {E.~S.}\ \bibnamefont
  {Battistelli}}, \bibinfo {author} {\bibfnamefont {F.}~\bibnamefont
  {Bellini}}, \bibinfo {author} {\bibfnamefont {C.}~\bibnamefont {Bucci}},
  \bibinfo {author} {\bibfnamefont {M.}~\bibnamefont {Calvo}}, \bibinfo
  {author} {\bibfnamefont {L.}~\bibnamefont {Cardani}}, \bibinfo {author}
  {\bibfnamefont {N.}~\bibnamefont {Casali}}, \bibinfo {author} {\bibfnamefont
  {M.}~\bibnamefont {Castellano}}, \bibinfo {author} {\bibfnamefont
  {I.}~\bibnamefont {Colantoni}}, \bibinfo {author} {\bibfnamefont
  {A.}~\bibnamefont {Coppolecchia}}, \bibinfo {author} {\bibfnamefont
  {C.}~\bibnamefont {Cosmelli}},  \emph {et~al.},\ }\href {\doibase
  https://doi.org/10.1063/1.4974082} {\bibfield  {journal} {\bibinfo  {journal}
  {The European Physical Journal C}\ }\textbf {\bibinfo {volume} {75}},\
  \bibinfo {pages} {353} (\bibinfo {year} {2015})}\BibitemShut {NoStop}%
\bibitem [{\citenamefont {Colantoni}\ \emph {et~al.}(2020)\citenamefont
  {Colantoni}, \citenamefont {Bellenghi}, \citenamefont {Calvo}, \citenamefont
  {Camattari}, \citenamefont {Cardani}, \citenamefont {Casali}, \citenamefont
  {Cruciani}, \citenamefont {Di~Domizio}, \citenamefont {Goupy}, \citenamefont
  {Guidi} \emph {et~al.}}]{BULLKID}%
  \BibitemOpen
  \bibfield  {author} {\bibinfo {author} {\bibfnamefont {I.}~\bibnamefont
  {Colantoni}}, \bibinfo {author} {\bibfnamefont {C.}~\bibnamefont
  {Bellenghi}}, \bibinfo {author} {\bibfnamefont {M.}~\bibnamefont {Calvo}},
  \bibinfo {author} {\bibfnamefont {R.}~\bibnamefont {Camattari}}, \bibinfo
  {author} {\bibfnamefont {L.}~\bibnamefont {Cardani}}, \bibinfo {author}
  {\bibfnamefont {N.}~\bibnamefont {Casali}}, \bibinfo {author} {\bibfnamefont
  {A.}~\bibnamefont {Cruciani}}, \bibinfo {author} {\bibfnamefont
  {S.}~\bibnamefont {Di~Domizio}}, \bibinfo {author} {\bibfnamefont
  {J.}~\bibnamefont {Goupy}}, \bibinfo {author} {\bibfnamefont
  {V.}~\bibnamefont {Guidi}},  \emph {et~al.},\ }\href {\doibase
  https://doi.org/10.1007/s10909-020-02408-3} {\bibfield  {journal} {\bibinfo
  {journal} {Journal of Low Temperature Physics}\ }\textbf {\bibinfo {volume}
  {199}},\ \bibinfo {pages} {593} (\bibinfo {year} {2020})}\BibitemShut
  {NoStop}%
\bibitem [{\citenamefont {Cruciani}\ \emph {et~al.}(2022)\citenamefont
  {Cruciani}, \citenamefont {Bandiera}, \citenamefont {Calvo}, \citenamefont
  {Casali}, \citenamefont {Colantoni}, \citenamefont {Del~Castello},
  \citenamefont {del Gallo~Roccagiovine}, \citenamefont {Delicato},
  \citenamefont {Giammei}, \citenamefont {Guidi}, \citenamefont {Goupy},
  \citenamefont {Pettinacci}, \citenamefont {Pettinari}, \citenamefont
  {Romagnoni}, \citenamefont {Tamisari}, \citenamefont {Mazzolari},
  \citenamefont {Monfardini},\ and\ \citenamefont {Vignati}}]{BULLKIDARRAY}%
  \BibitemOpen
  \bibfield  {author} {\bibinfo {author} {\bibfnamefont {A.}~\bibnamefont
  {Cruciani}}, \bibinfo {author} {\bibfnamefont {L.}~\bibnamefont {Bandiera}},
  \bibinfo {author} {\bibfnamefont {M.}~\bibnamefont {Calvo}}, \bibinfo
  {author} {\bibfnamefont {N.}~\bibnamefont {Casali}}, \bibinfo {author}
  {\bibfnamefont {I.}~\bibnamefont {Colantoni}}, \bibinfo {author}
  {\bibfnamefont {G.}~\bibnamefont {Del~Castello}}, \bibinfo {author}
  {\bibfnamefont {M.}~\bibnamefont {del Gallo~Roccagiovine}}, \bibinfo {author}
  {\bibfnamefont {D.}~\bibnamefont {Delicato}}, \bibinfo {author}
  {\bibfnamefont {M.}~\bibnamefont {Giammei}}, \bibinfo {author} {\bibfnamefont
  {V.}~\bibnamefont {Guidi}}, \bibinfo {author} {\bibfnamefont
  {J.}~\bibnamefont {Goupy}}, \bibinfo {author} {\bibfnamefont
  {V.}~\bibnamefont {Pettinacci}}, \bibinfo {author} {\bibfnamefont
  {G.}~\bibnamefont {Pettinari}}, \bibinfo {author} {\bibfnamefont
  {M.}~\bibnamefont {Romagnoni}}, \bibinfo {author} {\bibfnamefont
  {M.}~\bibnamefont {Tamisari}}, \bibinfo {author} {\bibfnamefont
  {A.}~\bibnamefont {Mazzolari}}, \bibinfo {author} {\bibfnamefont
  {A.}~\bibnamefont {Monfardini}}, \ and\ \bibinfo {author} {\bibfnamefont
  {M.}~\bibnamefont {Vignati}},\ }\href {\doibase 10.1063/5.0128723} {\bibfield
   {journal} {\bibinfo  {journal} {Applied Physics Letters}\ }\textbf {\bibinfo
  {volume} {121}},\ \bibinfo {pages} {213504} (\bibinfo {year} {2022})},\
  \Eprint
  {http://arxiv.org/abs/https://pubs.aip.org/aip/apl/article-pdf/doi/10.1063/5.0128723/16487869/213504\_1\_online.pdf}
  {https://pubs.aip.org/aip/apl/article-pdf/doi/10.1063/5.0128723/16487869/213504\_1\_online.pdf}
  \BibitemShut {NoStop}%
\bibitem [{\citenamefont {Sletten}\ \emph {et~al.}(2019)\citenamefont
  {Sletten}, \citenamefont {Moores}, \citenamefont {Viennot},\ and\
  \citenamefont {Lehnert}}]{QUBITPHONONS}%
  \BibitemOpen
  \bibfield  {author} {\bibinfo {author} {\bibfnamefont {L.~R.}\ \bibnamefont
  {Sletten}}, \bibinfo {author} {\bibfnamefont {B.~A.}\ \bibnamefont {Moores}},
  \bibinfo {author} {\bibfnamefont {J.~J.}\ \bibnamefont {Viennot}}, \ and\
  \bibinfo {author} {\bibfnamefont {K.~W.}\ \bibnamefont {Lehnert}},\ }\href
  {\doibase https://doi.org/10.1103/PhysRevX.9.021056} {\bibfield  {journal}
  {\bibinfo  {journal} {Phys. Rev. X}\ }\textbf {\bibinfo {volume} {9}},\
  \bibinfo {pages} {021056} (\bibinfo {year} {2019})}\BibitemShut {NoStop}%
\bibitem [{\citenamefont {Kittel}\ and\ \citenamefont {McEuen}(2018)}]{KITTEL}%
  \BibitemOpen
  \bibfield  {author} {\bibinfo {author} {\bibfnamefont {C.}~\bibnamefont
  {Kittel}}\ and\ \bibinfo {author} {\bibfnamefont {P.}~\bibnamefont
  {McEuen}},\ }\href@noop {} {\emph {\bibinfo {title} {Introduction to solid
  state physics}}}\ (\bibinfo  {publisher} {John Wiley \& Sons},\ \bibinfo
  {year} {2018})\BibitemShut {NoStop}%
\end{thebibliography}%
 \end{document}